\documentclass[prd,showkeys,floatfix,twocolumn,amsmath,amssymb]{revtex4}
\usepackage{graphicx}
\usepackage{multirow}
\usepackage{subfigure}
\usepackage[colorlinks,citecolor=blue,anchorcolor=red,menucolor=red,linkcolor=red,filecolor=red,runcolor=red,urlcolor=blue,frenchlinks=red]{hyperref}
\usepackage{enumitem}
\usepackage{slashed}
\usepackage[normalem]{ulem}
\usepackage{xcolor}
\usepackage{cleveref}

\makeatletter
\renewcommand{\@thesubfigure}{\hskip\subfiglabelskip}
\makeatother
\allowdisplaybreaks[3]

\begin{document}
\title{QCD Sum Rule Analysis of Triply Heavy $(Q\bar{Q})(Q\bar{q})$ Tetraquark States with $J^P=0^{\pm}$}
\author{Wen-Shuai Zhang$^{}$}
\author{Chun-Gui Duan$^{}$}
\email{duancg@hebtu.edu.cn}
\author{Zhi-Hui Guo$^{}$}
\email{zhguo@hebtu.edu.cn}
\author{Liang Tang$^{}$}
\email{tangl@hebtu.edu.cn}

\affiliation{College of Physics and Hebei Key Laboratory of Photophysics Research and Application, Hebei Normal University, Shijiazhuang 050024, China}

\begin{abstract}
Within the framework of QCD sum rules, we systematically investigate the mass spectra and possible decay patterns of the $(c\bar{c})(c\bar{q})$ and $(b\bar{b})(b\bar{q})$ tetraquark states with quantum numbers $J^{P}=0^{\pm}$. Based on two distinct color configurations, $[8_c]_{Q\bar{Q}} \otimes [8_c]_{Q\bar{q}}$ and $[1_c]_{Q\bar{Q}} \otimes [1_c]_{Q\bar{q}}$, we construct 18 interpolating currents for these states, and obtain stable sum rules for a subset of them. By calculating the corresponding two-point correlation functions, we extract their mass spectra. For the $(c\bar{c})(c\bar{q})$ system, we identify four possible tetraquark states: two with $J^{P}=0^+$, namely $T_{3c,0}(4760)$ and $T_{3c,0}(5000)$, and two with $J^{P}=0^-$, denoted as $T_{3c,0}(5040)$ and $T_{3c,0}(5370)$. For the $(b\bar{b})(b\bar{q})$ system, the extracted masses are found to lie in the ranges $13.72$--$14.02$ GeV for the $J^{P}=0^+$ states and $13.90$--$14.22$ GeV for the $J^{P}=0^-$ states. We further analyze their possible decay modes. Our results indicate that $T_{3c,0}(5000)$, $T_{3c,0}(5040)$, and $T_{3c,0}(5370)$ can decay into a charmonium state and a charmed meson, and are therefore expected to have appreciable decay widths. By contrast, $T_{3c,0}(4760)$ and all predicted $(b\bar{b})(b\bar{q})$ tetraquark states are expected to be relatively narrow, since the corresponding two-body strong decays via the fall-apart mechanism are kinematically forbidden.  Therefore, $T_{3c,0}(4760)$ and all predicted $(b\bar{b})(b\bar{q})$ tetraquark states are promising candidates for experimental searches in final states containing a $D$ or a $\bar{B}$ meson, accompanied by light hadrons or a photon.
\end{abstract}


\maketitle

\section{Introduction}
\label{sec:introduction}

In 1964, Gell-Mann and Zweig independently proposed the quark model for hadron classification, in which conventional hadrons are described as $q\bar{q}$ mesons or $qqq$ baryons, while more general multiquark configurations are not excluded~\cite{Gell-Mann,Zweig:1964ruk}. Within Quantum Chromodynamics (QCD), exotic hadronic states such as multiquark states, hybrid states, and glueballs are also allowed. The search for exotic hadrons has therefore become one of the most active topics in hadron physics, as it provides valuable insights into the nonperturbative dynamics of QCD. Since the discovery of the $X(3872)$ in 2003~\cite{Choi:2003(3872)}, an increasing number of XYZ states have been reported in experiments~\cite{pdg,Guo:2017jvc,Liu:2019zoy,xing2022review,gang2025review,Liu:2024uxn}. Most of these newly observed candidates contain at least one heavy quark/antiquark. The heavy exotic hadron candidates in the singly heavy, doubly heavy, and fully heavy tetraquark sectors have been experimentally established, while triply heavy tetraquark states are not yet observed and have attracted growing theoretical interests.  

The observation of the narrow resonance $D_{s0}(2317)$ by the BaBar collaboration in 2003 in the $D_s^+\pi^0$ channel~\cite{BaBar:2317(2003)} marked a turning point in the study of singly heavy exotic states. This state, together with $D_{s1}(2460)$ subsequently observed by CLEO~\cite{CLEO2460and2317(2003)} and later confirmed by Belle~\cite{Belle2317(2003)}, have masses significantly below the expectations of the conventional quark model~\cite{quarkmodel-cs}. Although some theoretical works attempted to interpret them as $P$-wave charmed-strange mesons~\cite{Cahn:2003-Pwave-cs,p-wave-cs-2}, the mass discrepancies between experimental measurements and quark model predictions have triggered a longstanding debate on their internal structures, with proposals of the $DK$ or $D^{*}K$ molecular configurations~\cite{molecular-DK-1,molecular-DK-2,molecular-DK-3,molecular-DK-4,molecular-DK-5,molecular-DK-6,molecular-DK-7,molecular-DK-8,molecular-DK-9} and compact tetraquark states~\cite{compact-DK-1,compact-DK-2,compact-DK-3,compact-DK-4,compact-DK-5,compact-DK-6}.

Among exotic candidates containing two heavy quarks/antiquarks or a heavy quark-antiquark pair, the $X(3872)$ is the most extensively studied example. First discovered by the Belle collaboration in 2003 in the decay $B^{\pm}\to K^{\pm}\pi^{+}\pi^{-}J/\psi$~\cite{Choi:2003(3872)} and subsequently confirmed by several experiments~\cite{CMS:2020-3872,BESIII:2019-3872-1,BESIII:2019-3872-2,BESIII:2020-3872}, its mass lies remarkably close to the $D\bar{D}^{*}$ threshold. This proximity has led to a variety of theoretical interpretations, such as the $\chi_{c1}(2P)$ charmonium assignment~\cite{charmonium2p-3872-1,charmonium2p-3872-2,charmonium2p-3872-3}, the hadronic molecular states of the $D^{(*)}\bar{D}^{(*)}$ systems~\cite{molecular-3872-DD}, and the $\bar{D}_s^{(*)} D_s^{(*)}$ bound states as hidden-strange partners~\cite{molecular-3872-DsDs}. Another milestone was the discovery of the charged $Z_c(3900)^{\pm}$ simultaneously by BESIII~\cite{BESIII:zc3900+-} and Belle~\cite{Belle:zc3900+-} in the $J/\psi\pi^{\pm}$ invariant mass spectrum of the $e^+e^- \to J/\psi\pi^+\pi^-$ reaction, and further indications of a neutral structure in the same mass region were later discussed based on the CLEO-c data~\cite{CLEO-c:zc3900-0}. Theoretical studies have interpreted $Z_c(3900)$ as a compact tetraquark~\cite{compact-3900-1,compact-3900-2} or as a $D\bar{D}^*$ resonant or virtual molecular state~\cite{molecular-3900-1,molecular-3900-2} or a mixture of both types~\cite{Yan:2023bwt,Sadl:2024dbd,Chen:2026fnz}. A series of additional doubly heavy candidates were  reported as well, including $Z_c(4020)$~\cite{zc4020-1}, $X(4050)$~\cite{x4050-1}, $X(4055)$~\cite{x4055-1}, $X(4100)$~\cite{x4100-1}, $Z_c(4200)$~\cite{zc4200-1}, $Z_c(4430)$~\cite{zc4430-1}, and the bottom counterparts $Z_b(10610)$ and $Z_b(10650)$~\cite{zb10610andzb10650-1}, all of which have been considered as possible multiquark states or hadronic molecules.

Another breakthrough came in 2020, when the LHCb collaboration observed the structure $X(6900)$ in the $J/\psi J/\psi$ mass spectrum~\cite{x6900-1}, providing the first compelling candidate for a fully charmed tetraquark system with quark content $c\bar{c}c\bar{c}$. This observation was soon followed by the report of additional structures, $X(6600)$ and $X(7200)$, by the CMS~\cite{CMS:x6600-7200-1} and ATLAS~\cite{ATLAS:x6600-7200-1} collaborations. These experimental findings have stimulated extensive theoretical activity, with proposed interpretations including compact tetraquark states~\cite{x6900-tetraquark-1,x6900-tetraquark-2,Guo:2020pvt,Kuang:2023vac,Liang:2021fzr} and hybrid states~\cite{x6900-hybrid-1,x6900-hybrid-2}, thereby offering new opportunities to probe the nonperturbative dynamics of heavy quarks in QCD.

To date, there have been no experimental reports of triply heavy tetraquark states yet. Nevertheless, substantial theoretical progress has been made in recent years. A variety of methods have been employed to explore the properties of triply heavy tetraquark systems, including the MIT bag model~\cite{triply:MIT-big}, effective field theory~\cite{triply:EFT}, the extended relativistic quark model~\cite{triply:quarkmodel-1}, the extended color-magnetic model~\cite{triply:chromomagnetic-model}, Regge trajectories~\cite{triply:Regge-trajectories}, QCD sum rules~\cite{triply:QCDsumrules-1,triply:QCDsumrules-2}, and the constituent quark model~\cite{triply:constituent-quark-model-1,triply:constituent-quark-model-2,triply:constituent-quark-model-3}. Using the extended relativistic quark model, Ref.~\cite{triply:quarkmodel-1} found that the masses of all investigated triply heavy tetraquark states lie above the corresponding meson-meson thresholds, and no bound states below those thresholds were obtained. In contrast, Ref.~\cite{triply:QCDsumrules-1} proposed the existence of resonance candidates with quantum numbers $J^P=0^+$ and $J^P=1^+$ below the corresponding bottomonium-$\bar{B}^{(*)}$ thresholds, which may therefore be relatively narrow.

In this work, within the framework of QCD sum rules, we systematically study the $(Q\bar{Q})(Q\bar{q})$ tetraquark states with quantum numbers $J^P=0^{\pm}$. We construct ten interpolating currents for the $J^P=0^+$ channel and eight interpolating currents for the $J^P=0^-$ channel, and then calculate the corresponding two-point correlation functions and spectral densities. In the numerical analysis, reliable sum rules are required to satisfy the following three criteria: good convergence of the operator product expansion (OPE), sufficient pole dominance over the continuum contribution, and stability of the extracted mass against variations of the Borel parameter. After imposing these criteria, we obtain stable sum rules for several currents in the $(Q\bar{Q})(Q\bar{q})$ system with $J^P=0^{\pm}$ and extract the corresponding mass spectra.

The remainder of this paper is organized as follows. In Section~\ref{sec:current}, we explicitly construct the interpolating currents for the $(Q\bar{Q})(Q\bar{q})$ configuration with quantum numbers $J^{P}=0^{\pm}$. In Section~\ref{sec:sumrule}, we present the QCD sum rule formalism for these currents. In Section~\ref{sec:numerical}, we give the numerical analysis and phenomenological discussions. Finally, the conclusions are summarized in Section~\ref{sec:summary}.

\section{Interpolating currents}
\label{sec:current}

In this work, we investigate the \((Q\bar{Q})(Q\bar{q})\) tetraquark system with quantum numbers \(J^{P}=0^{\pm}\) using local tetraquark interpolating currents. For the two color structures \([8_c]_{Q\bar{Q}} \otimes [8_c]_{Q\bar{q}}\) and \([1_c]_{Q\bar{Q}} \otimes [1_c]_{Q\bar{q}}\), we construct all local four-quark currents without covariant derivatives, systematically taking into account all allowed Dirac structures. For the scalar channel (\(J^{P}=0^{+}\)), we construct five color-octet--color-octet currents,
\begin{eqnarray}
&J_1 =& (\bar{Q}^a \lambda^n_{ab} Q^b)(\bar{q}^c \lambda^n_{cd} Q^d), \label{eq:liu6}\\
&J_{2} =& (\bar{Q}^a \gamma_5 \lambda^n_{ab} Q^b)(\bar{q}^c \gamma_5 \lambda^n_{cd} Q^d),\label{eq:liu10} \\
&J_{3} =& (\bar{Q}^a \sigma_{\mu\nu} \lambda^n_{ab} Q^b)(\bar{q}^c \sigma^{\mu\nu} \lambda^n_{cd} Q^d), \label{eq:liu8}  \\
&J_{4}=& (\bar{Q}^a \gamma_\mu \lambda^n_{ab} Q^b)(\bar{q}^c \gamma^\mu \lambda^n_{cd} Q^d), \label{eq:liu7}\\
&J_{5} =& (\bar{Q}^a \gamma_\mu \gamma_5 \lambda^n_{ab} Q^b)(\bar{q}^c \gamma^\mu \gamma_5 \lambda^n_{cd} Q^d),\label{eq:liu9}
\end{eqnarray}
and five color-singlet--color-singlet currents,
\begin{eqnarray}
&J_{6} =& (\bar{Q}_a Q^a)(\bar{q}_b Q^b),\label{eq:liu1} \\
&J_{7} =& (\bar{Q}_a \gamma_5 Q^a)(\bar{q}_b \gamma_5 Q^b), \label{eq:liu5}\\
&J_8=& (\bar{Q}_a \sigma_{\mu\nu} Q^a)(\bar{q}_b \sigma^{\mu\nu} Q^b),\label{eq:liu3}   \\
&J_9 =& (\bar{Q}_a \gamma_\mu Q^a)(\bar{q}_b \gamma^\mu Q^b), \label{eq:liu2}  \\
&J_{10} =& (\bar{Q}_a \gamma_\mu \gamma_5 Q^a)(\bar{q}_b \gamma^\mu \gamma_5 Q^b). \label{eq:liu4}
\end{eqnarray}

For the pseudoscalar channel (\(J^{P}=0^{-}\)), we construct four color-octet--color-octet currents,
\begin{eqnarray}
&J_{11} =& (\bar{Q}^a \lambda^n_{ab} Q^b)(\bar{q}^c \gamma_5 \lambda^n_{cd} Q^d),\label{eq:liu13} \\
&J_{12} =& (\bar{Q}^a \gamma_5 \lambda^n_{ab} Q^b)(\bar{q}^c \lambda^n_{cd} Q^d),\label{eq:liu14}\\
&J_{13} =& (\bar{Q}^a \gamma_{\mu} \lambda^n_{ab} Q^b)(\bar{q}^c \gamma^{\mu} \gamma_5 \lambda^n_{cd} Q^d),\label{eq:liu17} \\
&J_{14} =& (\bar{Q}^a \gamma_{\mu} \gamma_5 \lambda^n_{ab} Q^b)(\bar{q}^c \gamma^{\mu} \lambda^n_{cd} Q^d),\label{eq:liu18}
\end{eqnarray}
and four color-singlet--color-singlet currents,
\begin{eqnarray}
&J_{15} =& (\bar{Q}_a Q^a)(\bar{q}_b \gamma_5 Q^b), \label{eq:liu11}\\
&J_{16} =& (\bar{Q}_a \gamma_5 Q^a)(\bar{q}_b Q^b), \label{eq:liu12}\\
&J_{17} =& (\bar{Q}_a \gamma_{\mu} Q^a)(\bar{q}_b \gamma^{\mu} \gamma_5 Q^b), \label{eq:liu15}\\
&J_{18} =& (\bar{Q}_a \gamma_{\mu} \gamma_5 Q^a)(\bar{q}_b \gamma^{\mu} Q^b). \label{eq:liu16}
\end{eqnarray}
Here, \(q\) denotes the light-quark field (\(u\) or \(d\)), \(Q\) denotes the heavy-quark field (\(c\) or \(b\)), and \(a,b,c,d\) are color indices. The matrices \(\lambda^n\) (\(n=1,2,\ldots,8\)) are the Gell-Mann SU(3) matrices, and the repeated adjoint index \(n\) is implicitly summed. The Lorentz indices in the above currents are understood to be fully contracted, so that the resulting operators carry the quantum numbers \(J^{P}=0^{\pm}\).

The above operators constitute the set of local bilinear-bilinear currents considered in the present work. Although currents with different color structures can be rearranged into other operator bases through Fierz transformations, they may couple to the same hadronic state with different strengths, and it is therefore meaningful to analyze them separately in the QCD sum rule approach~\cite{Chen:2007xr,Chen:2016qju}.

In the subsequent numerical analysis, we will examine these currents according to the three standard criteria introduced in Section~\ref{sec:introduction}: OPE convergence, pole dominance, and Borel stability. We find that only \(J_3\), \(J_4\), \(J_7\), \(J_{10}\), \(J_{15}\), and \(J_{18}\) can lead to stable and reliable sum rules. Therefore, we will only focus on the phenomenological discussions for these six currents in the following parts.


\section{QCD sum rule analysis}
\label{sec:sumrule}

In this section, we study the \((Q\bar{Q})(Q\bar{q})\) tetraquark states within the framework of QCD sum rules. For a scalar or pseudoscalar interpolating current \(J(x)\), we consider the two-point correlation function
\begin{equation}
\Pi(q^2) = i \int d^4x\, e^{iq \cdot x} \langle 0 | T[J(x) J^\dagger(0)] | 0 \rangle .
\label{eq:correlation}
\end{equation}
At the hadronic level, the correlation function satisfies the following dispersion relation
\begin{equation}
\Pi(q^2)=\int_{s_<}^{\infty}\frac{\rho(s)}{s-q^2-i\epsilon}\,ds+\text{subtraction terms},
\label{eq:dispersion}
\end{equation}
where \(\rho(s)\equiv \mathrm{Im}\,\Pi(s)/\pi\) is the spectral density. In the present work, under the approximation \(m_q=0\), the lower limit can be taken as \(s_< \simeq 9m_Q^2\). The subtraction terms, being polynomial in \(q^2\), are removed by the Borel transformation.

By inserting a complete set of hadronic intermediate states, the phenomenological spectral density can be parameterized as
\begin{equation}
\begin{aligned}
\rho_{\text{phen}}(s)
&\equiv \sum_n \delta(s-M_n^2)\langle 0|J|n\rangle \langle n|J^\dagger|0\rangle \\
&= f_X^2 \delta(s-M_X^2) + \theta(s-s_0)\rho^{\rm cont}(s) \, ,
\end{aligned}
\label{eq:phen}
\end{equation}
where the first term denotes the contribution of the lowest-lying state \(|X;0^\pm\rangle\), and the second term represents the continuum contribution. \(M_X\) is the mass of the hadronic state, and the coupling \(f_X\) is defined as 
\begin{equation}
\langle 0|J|X\rangle = f_X \, .
\label{eq:coupling}
\end{equation}
For the dimension-6 tetraquark currents used in this work, \(f_X\) has mass dimension 5. The continuum threshold \(s_0\) separates the ground-state contribution from higher resonances and continuum states, and the spectral density of the latter two objects is approximated by the OPE result under the quark-hadron duality ansatz.

At the quark-gluon level, the correlation function in Eq.~(\ref{eq:correlation}) is evaluated by means of the OPE. For the light \(u/d\) quark, we use the propagator in coordinate space,
\begin{equation}
\begin{aligned}
S_{ab}^q(x) &= \frac{i \delta_{ab} \slashed{x}}{2\pi^2 x^4}
- \frac{i t_{ab}^N G^{N}_{\alpha \beta}}{32 \pi^2 x^2}
\left( \sigma_{\alpha \beta} \slashed{x} + \slashed{x} \sigma_{\alpha \beta} \right) \\
&\quad - \frac{\delta_{ab}}{12} \langle \bar{q}q \rangle
- \frac{\delta_{ab} x^2}{192} \langle \bar{q}g_s \sigma G q \rangle \\
&\quad - \frac{t_{ab}^N \sigma_{\alpha \beta}}{192} \langle \bar{q}g_s \sigma G q \rangle ,
\label{eq:q propagator}
\end{aligned}
\end{equation}
while the heavy-quark propagator is conveniently written in momentum space as
\begin{widetext}
\begin{equation}
\begin{aligned}
S_{ab}^Q(x) &=  \int \frac{d^4 p}{(2\pi)^4} e^{-ip\cdot x}
\bigg\{ \frac{i\delta_{ab}(\slashed{p}+m_Q)}{p^2-m_Q^2}
 -  ig_s t^N_{ab} G^N_{\alpha \beta}
 \frac{(\slashed{p}+m_Q)\sigma^{\alpha \beta}+\sigma^{\alpha \beta}(\slashed{p}+m_Q)}{4(p^2-m_Q^2)^2}   \\
&\left. \quad - \frac{i}{4}g_s^2(t^N t^M)_{ab}G_{\mu\rho}^N G_{\nu\sigma}^M(\slashed{p}+m_Q)
 \frac{ (f^{\mu\rho\nu\sigma} + f^{\mu\nu\rho\sigma} + f^{\mu\nu\sigma\rho})}{(p^2 - m_Q^2)^5} \right. \\
&\left. \quad + \frac{g_s^3 (t^N  t^M t^L)_{ab} G^N_{\mu \alpha}G^M_{\nu \beta}G^L_{\rho \gamma}(\slashed{p}+m_Q)}{8(p^2-m_Q^2)^7}
(f^{\mu \alpha \nu \beta \rho \gamma}+f^{\mu \alpha \nu  \rho \beta \gamma}+f^{\mu \alpha \nu  \rho  \gamma \beta}
+f^{\mu \nu \alpha \beta \rho \gamma } \right.\\
&\left. \qquad +f^{\mu \nu \beta \alpha \rho \gamma }+f^{\mu \nu \beta \rho \alpha \gamma }
+f^{\mu \nu \beta \rho \gamma \alpha }+f^{\mu \nu \alpha \rho \beta \gamma }
+f^{\mu \nu \alpha \rho \gamma \beta }+f^{\mu \nu \rho \alpha \beta \gamma } \right.\\
& \qquad +f^{\mu \nu \rho \beta \alpha \gamma }+f^{\mu \nu \rho \beta \gamma \alpha }
+f^{\mu \nu \rho \alpha \gamma \beta }+f^{\mu \nu \rho \gamma \alpha \beta }
+f^{\mu \nu \rho \gamma \beta \alpha }) \bigg\} ,
\label{eq:Q propagator}
\end{aligned}
\end{equation}
\end{widetext}
where \(a\) and \(b\) are color indices. We have defined
\[
f^{\mu \nu \cdots \alpha \beta}
\equiv \gamma^{\mu}(\slashed{p} + m_Q)\gamma^{\nu}(\slashed{p} + m_Q)\cdots
\gamma^{\alpha}(\slashed{p} + m_Q)\gamma^{\beta}(\slashed{p} + m_Q) .
\]

To suppress the contributions of higher resonances and continuum states and, at the same time, improve the convergence of the OPE, we perform the Borel transformation 
\begin{equation}
\hat{B} \left[ f(q^2) \right]
= \lim_{\substack{-q^2, n \to \infty \\ -q^2 / n \equiv M_B^2}}
 \frac{1}{n!} \left(-q^2\right)^{n+1}
 \left( \frac{d}{dq^2} \right)^n f(q^2),
\label{eq:Borel transformation}
\end{equation}
where \(M_B^2\) is the Borel parameter. The polynomial subtraction terms in Eq.~(\ref{eq:dispersion}) vanish after the Borel transformation.

\begin{figure}[hbtp]
\begin{center}
\scalebox{0.15}{\includegraphics{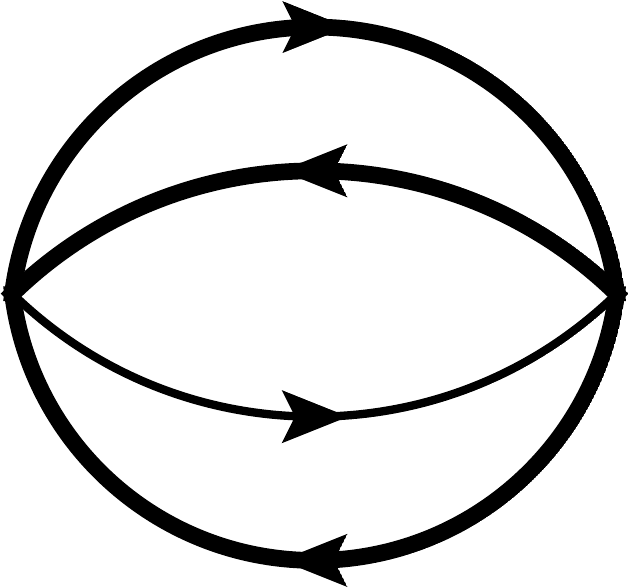}}~~
\\[2mm]
\scalebox{0.15}{\includegraphics{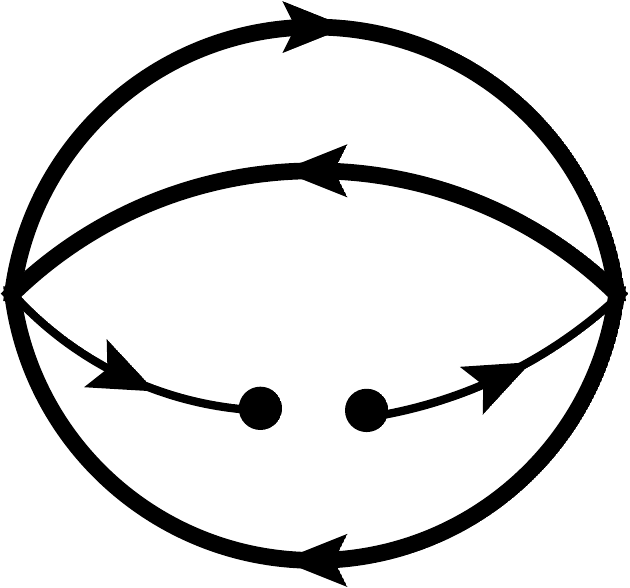}}~~
\scalebox{0.15}{\includegraphics{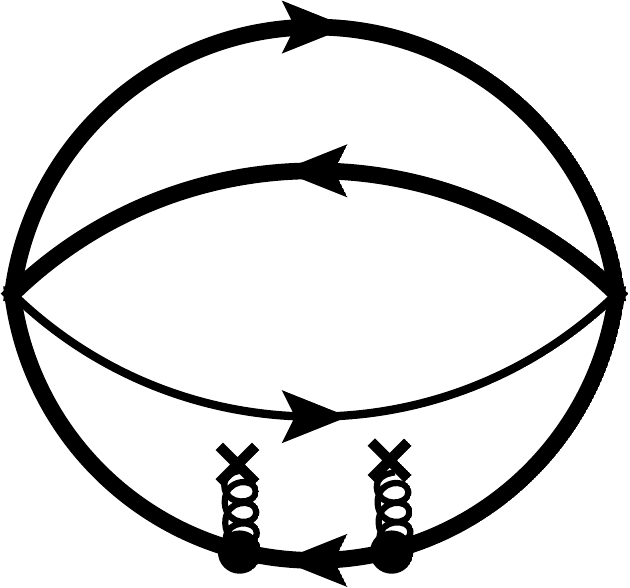}}~~
\\[2mm]
\scalebox{0.15}{\includegraphics{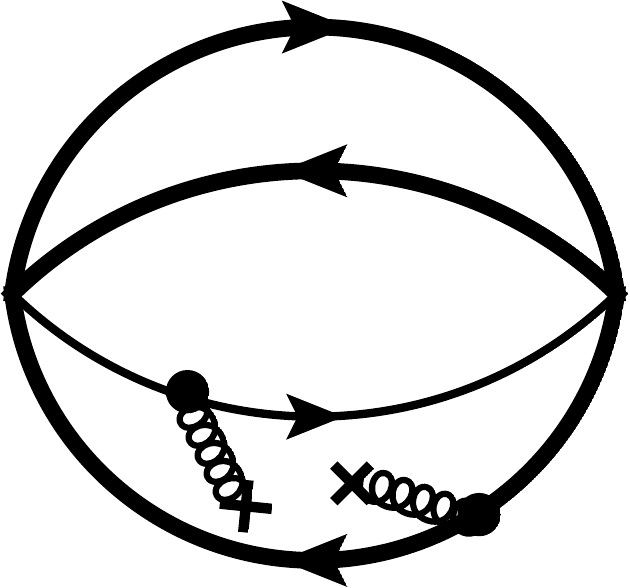}}~~
\scalebox{0.15}{\includegraphics{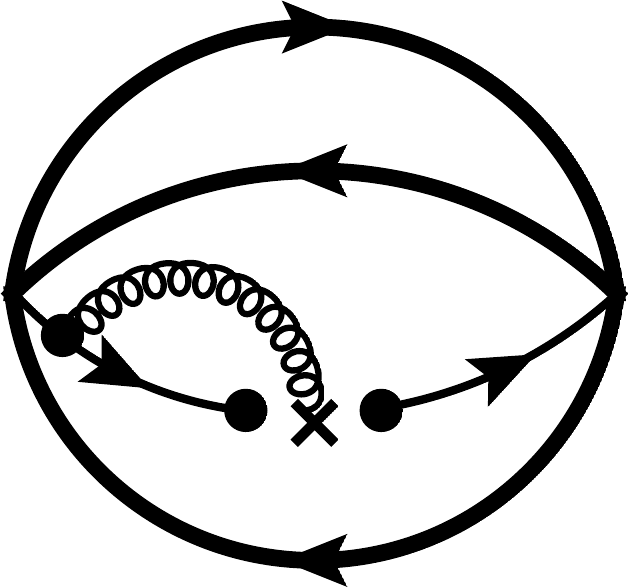}}~~
\scalebox{0.15}{\includegraphics{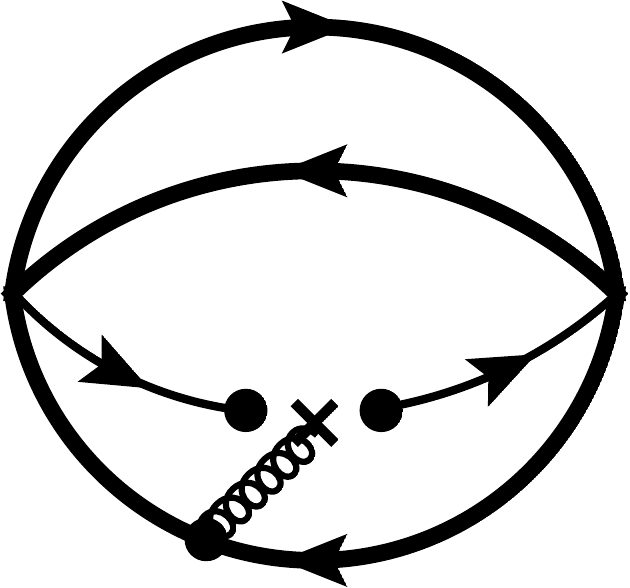}}~~
\\[2mm]
\scalebox{0.15}{\includegraphics{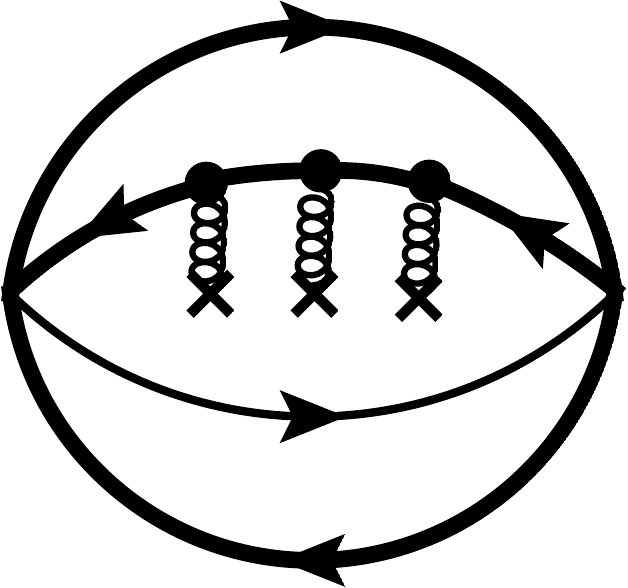}}~~
\scalebox{0.15}{\includegraphics{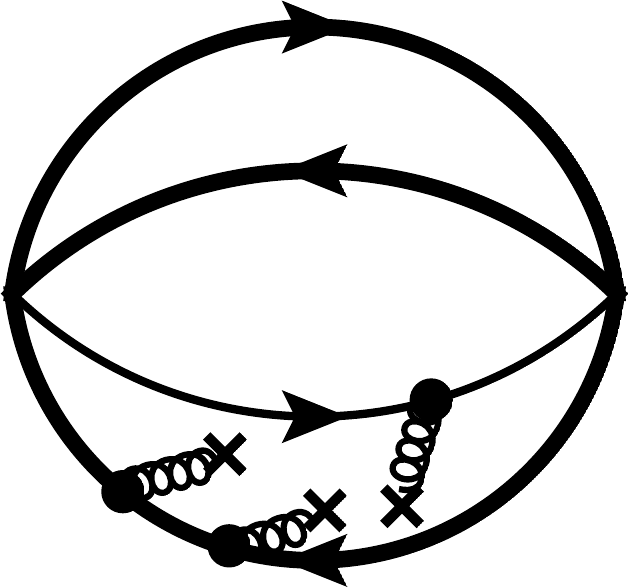}}~~
\scalebox{0.15}{\includegraphics{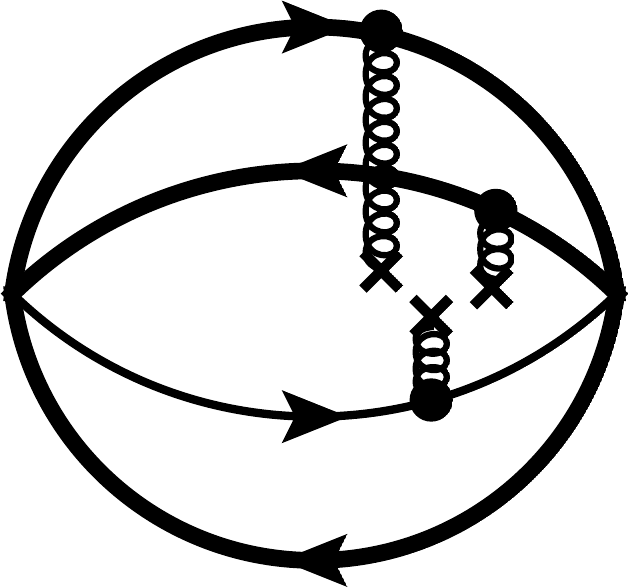}}~~
\scalebox{0.15}{\includegraphics{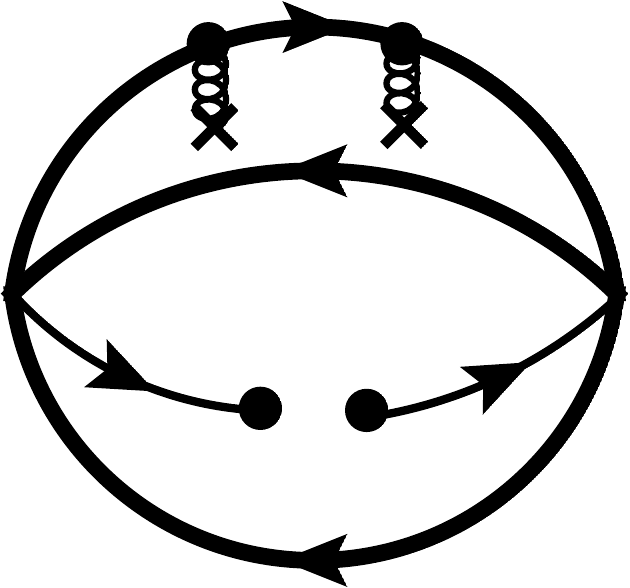}}~~
\\[2mm]
\scalebox{0.15}{\includegraphics{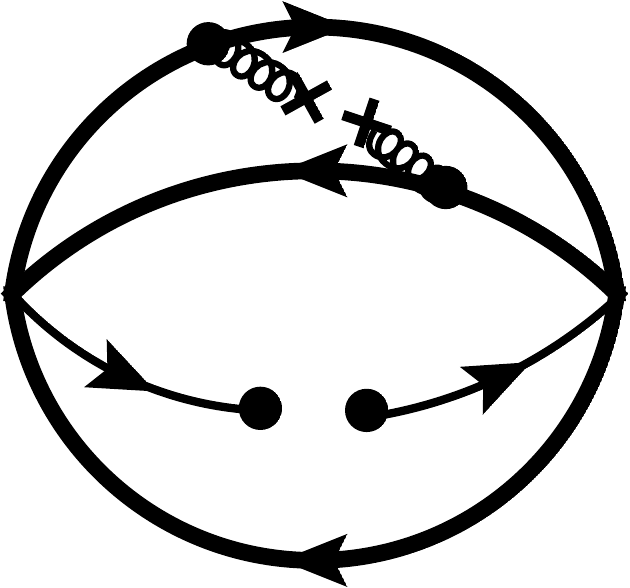}}~~
\scalebox{0.15}{\includegraphics{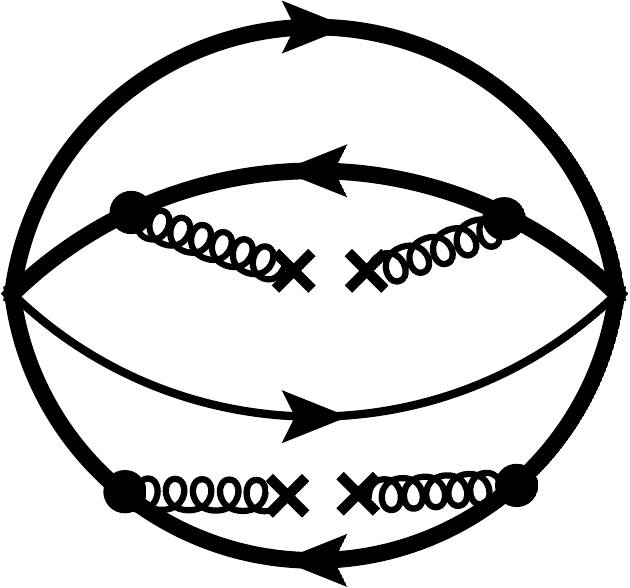}}~~
\scalebox{0.15}{\includegraphics{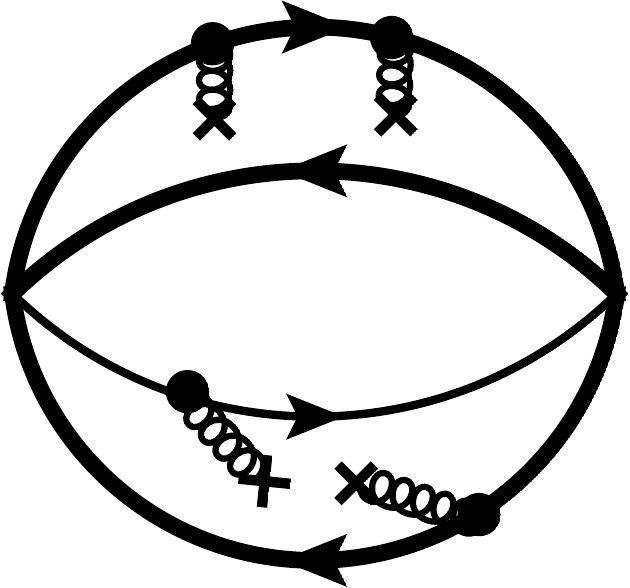}}~~
\scalebox{0.15}{\includegraphics{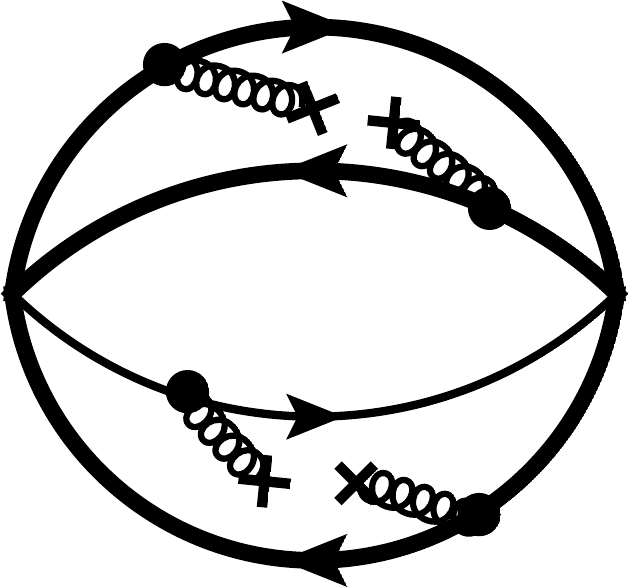}}~~
\\[2mm]
\scalebox{0.15}{\includegraphics{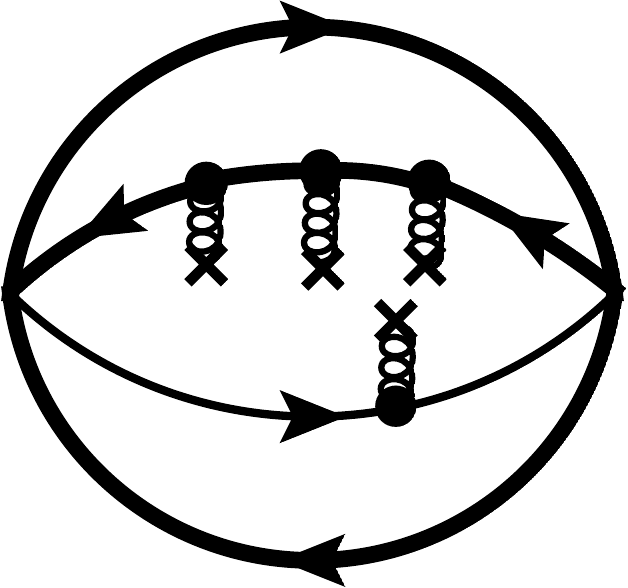}}~~
\scalebox{0.15}{\includegraphics{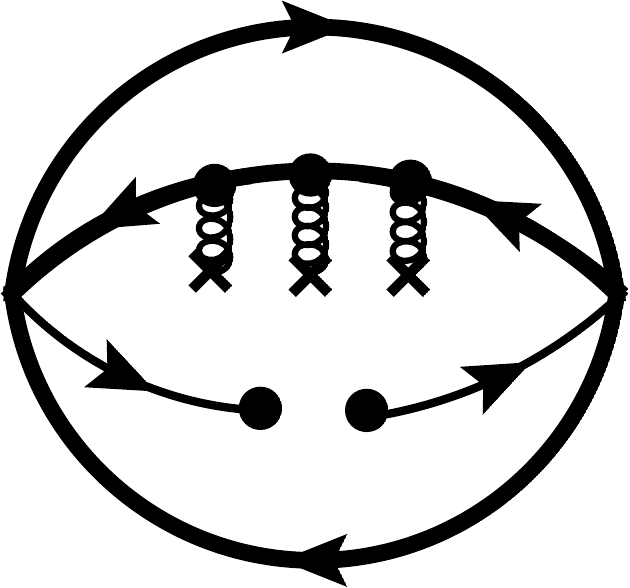}}~~
\scalebox{0.15}{\includegraphics{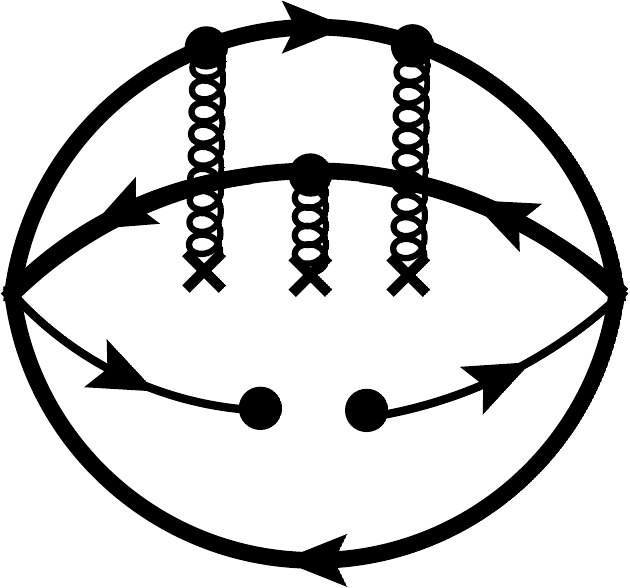}}~~
\scalebox{0.15}{\includegraphics{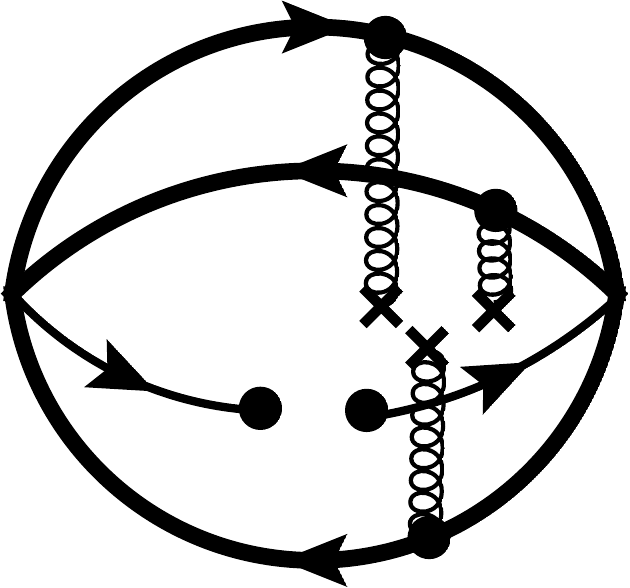}}~~
\\[2mm]
\scalebox{0.15}{\includegraphics{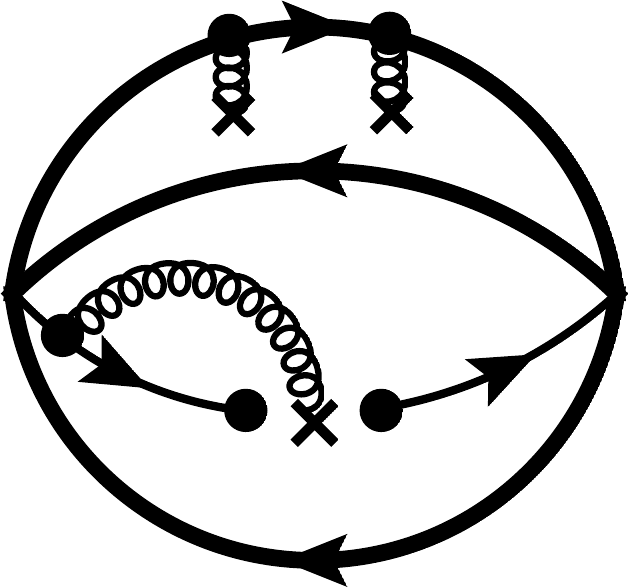}}~~
\scalebox{0.15}{\includegraphics{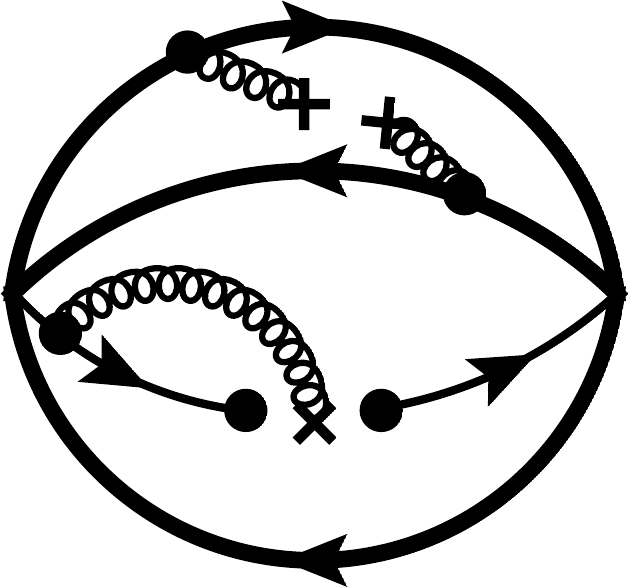}}~~
\scalebox{0.15}{\includegraphics{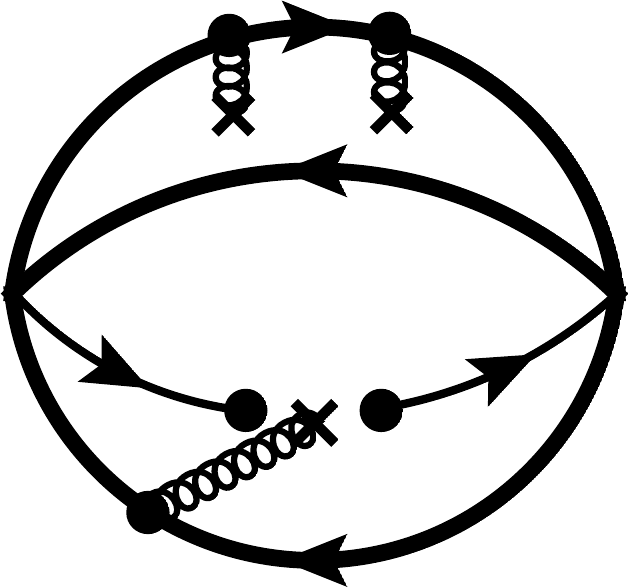}}~~
\scalebox{0.15}{\includegraphics{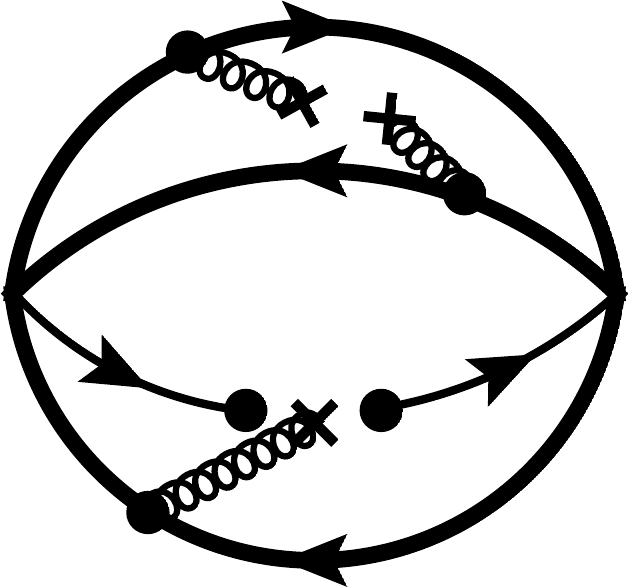}}~~
\caption{Relevant Feynman diagrams to the two-point correlation functions at the quark-hadron level. All possible permutations are implicitly included. }
\label{fig:feynman}
\end{center}
\end{figure}

At the quark-gluon level, we calculate the spectral density \(\rho_{\text{OPE}}(s)\), i.e, the counterpart of the hadronic one in Eq.~(\ref{eq:phen}), by expanding the correlation function in powers of local operators. The explicit expressions of the spectral densities are collected in Appendix~\ref{sec:appendix}. The corresponding Feynman diagrams are shown in Fig.~\ref{fig:feynman}. In this work, we keep the leading order \(\alpha_s\) correction and retain the OPE contributions up to dimension 9, including the perturbative term, the quark condensate \(\langle \bar{q}q\rangle\), the gluon condensate \(\langle \alpha_s GG\rangle\), the quark-gluon mixed condensate \(\langle g_s\bar{q}\sigma G q\rangle\), the three-gluon condensate \(\langle g_s^3 GGG\rangle\), the four-gluon condensate \(\langle G^4 \rangle\), and all the possible combinations among them. The four-gluon condensate is parameterized as \(\langle G^4\rangle \sim \kappa \langle GG\rangle^2\), where \(\kappa\) characterizes the possible violation of vacuum saturation~\cite{vacuum-saturation:1,vacuum-saturation:2}. The uncertainty associated with this approximation will be discussed in Section~\ref{sec:numerical}.

By matching the hadronic representation to the OPE side and applying the Borel transform, we obtain the sum rule
\begin{equation}
\begin{split}
\Pi(s_0,M_B^2)
&= f_X^2 e^{-M_X^2/M_B^2} \\
&= \int_{s_<}^{s_0} e^{-s/M_B^2}\rho_{\text{OPE}}(s,1/M_B^2)\, ds ,
\end{split}
\label{eq:sum rule}
\end{equation}
where the upper limit \(s_0\) is introduced under the quark-hadron duality approximation.

The hadron mass can then be extracted from
\begin{equation}
M_{X}^2 =
-\frac{\dfrac{\partial}{\partial\tau}\int_{s_<}^{s_0} ds\, \rho_{\text{OPE}}(s,\tau)\, e^{-\tau s}}
{\int_{s_<}^{s_0} ds\, \rho_{\text{OPE}}(s,\tau)\, e^{-\tau s}}
\bigg|_{\tau = \frac{1}{M_{B}^2}}\, .
\label{eq:mass}
\end{equation}
The coupling constant is given by 
\begin{eqnarray}
f^2_X(s_0,M_B^2)
= \left[\int^{s_0}_{s_<} \rho_{\rm OPE}(s)e^{-s/M_B^2}ds\right]
e^{M_X^2/M_B^2} .
\label{eq:decay}
\end{eqnarray}
The choices of the continuum threshold \(s_0\) and the Borel window \(M_B^2\) will be discussed in detail in the next section.

\section{Numerical analysis}
\label{sec:numerical}

For the numerical analysis within the QCD sum-rule framework, we use the following input parameters for the heavy-quark masses and vacuum condensates~\cite{pdg,condensates:Stephan-Narison-1,condensates:Stephan-Narison-2,condensates:quark-gluon-1,condensates:quark-quark-1}:
\begin{equation}
\begin{aligned}
    & m_q = m_u = m_d = 0, \\
    & m_c(m_c) = \overline{m}_c = (1.27 \pm 0.02)\,\text{GeV}, \\
    & m_b(m_b) = \overline{m}_b = 4.18 ^{+0.04}_{-0.03}\,\text{GeV}, \\
    & \langle \bar{q} q \rangle = - (0.240 \pm 0.010)^3 \, \text{GeV}^3, \\
    & \langle \bar{q} g_s \sigma G q \rangle = (0.8 \pm 0.2)\,\text{GeV}^2 \times \langle \bar{q} q \rangle, \\
    & \langle \alpha_s G G \rangle = (6.35 \pm 0.35) \times 10^{-2} \, \text{GeV}^4, \\
    & \langle g_s^3 G^3 \rangle = (8.2 \pm 1.0)\,\text{GeV}^2 \times \langle \alpha_s G G \rangle .
\end{aligned}
\label{eq:condensate}
\end{equation}
The heavy-quark masses at the corresponding energy scales are obtained  using two-loop perturbative QCD in the $\overline{\rm MS}$ scheme~\cite{pdg}. 

As can be seen from Eq.~(\ref{eq:mass}), the hadron mass depends on two auxiliary parameters: the Borel parameter \(M_B^2\) and the continuum threshold \(s_0\). To determine their working regions, we adopt the following criteria: (i) OPE should show good convergence; (ii) the pole contribution should be sufficiently large; (iii) the extracted mass should exhibit only weak dependence on \(M_B^2\) and \(s_0\).

To ensure the convergence of the OPE, we require the higher dimensional contributions to remain sufficiently small. According to Ref.~\cite{suniu:1}, we adopt the following rules 
\begin{eqnarray}
\mbox{CVG}_A &\equiv& \left|\frac{ \Pi^{{\rm D=9+8}}(\infty, M_B^2) }{ \Pi(\infty, M_B^2) }\right| \leq 5\% \, , \label{eq:CVG_A}
\\
\mbox{CVG}_B &\equiv& \left|\frac{ \Pi^{{\rm D=7+6}}(\infty, M_B^2) }{ \Pi(\infty, M_B^2) }\right| \leq 10\% \, , \label{eq:CVG_B}
\\
\mbox{CVG}_C &\equiv& \left|\frac{ \Pi^{{\rm D=5+4}}(\infty, M_B^2) }{ \Pi(\infty, M_B^2) }\right| \leq 20\% \, , \label{eq:CVG_C}
\end{eqnarray}
where the denominator denotes the full OPE result up to dimension 9, and the numerators represent the contributions from the dimensions indicated in the superscripts. 

\begin{figure}[hbt]
\begin{center}
\includegraphics[width=0.445\textwidth]{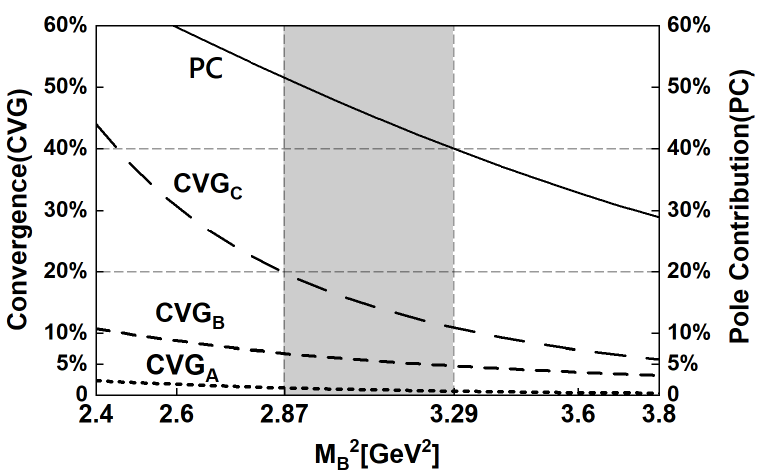}
\caption{The quantities CVG$_{A/B/C}$ and the pole contribution PC as functions of \(M_B^2\) for the current \(J_3\), with \(s_0=29.0~\text{GeV}^2\).}
\label{fig:cvgpole}
\end{center}
\end{figure}

We take the current \(J_3\) as an example to illustrate how the mass and current coupling of the corresponding hadronic state are extracted. As shown by the three types of dashed curves in Fig.~\ref{fig:cvgpole}, the criterion \(\mathrm{CVG}_C\) in Eq.~(\ref{eq:CVG_C}) requires the Borel parameter to satisfy $M_B^2 \geq 2.87~\text{GeV}^2$. 
In this case, the \(\mathrm{CVG}_A\) and \(\mathrm{CVG}_B\) criteria are automatically satisfied once \(\mathrm{CVG}_C\) is fulfilled.

To ensure a sufficiently large pole contribution (PC), we further require
\begin{equation}
\mbox{PC} \equiv \left|\frac{ \Pi(s_0, M_B^2) }{ \Pi(\infty, M_B^2) }\right| \geq 40\% \, ,
\label{eq:pole_contribution_cond}
\end{equation}
where \(\Pi(s_0,M_B^2)\) denotes the Borel-transformed correlation function integrated up to the continuum threshold \(s_0\). As shown by the solid curve in Fig.~\ref{fig:cvgpole}, for \(s_0 = 29.0~\text{GeV}^2\), the condition in Eq.~(\ref{eq:pole_contribution_cond}) leads to the upper bound $M_B^2 \leq 3.29~\text{GeV}^2$. 
For each value of \(s_0\), there is a corresponding upper bound on \(M_B^2\). We also find that a reasonable Borel window exists only when \(s_0\) is larger than the minimum value of $s_0^{\rm min} = 26.7~\text{GeV}^2$. 

\begin{figure*}[hbtp]
\begin{center}
\subfigure[]{\includegraphics[width=0.35\textwidth]{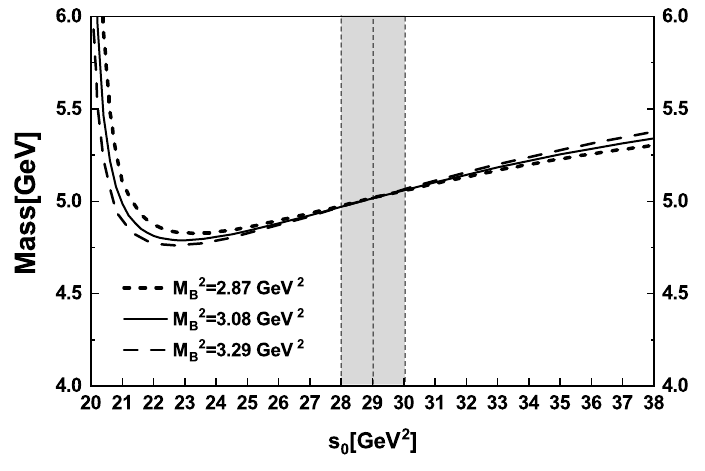}}
~~~~~~~~~~
\subfigure[]{\includegraphics[width=0.35\textwidth]{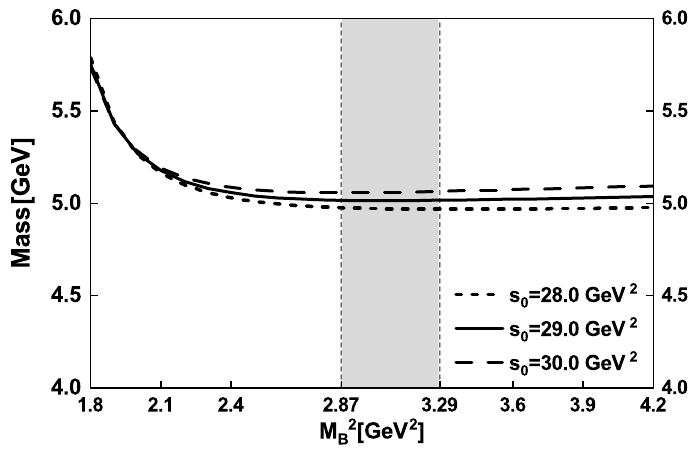}}
\caption{The mass \(M_{0^+}^{J_3}\), extracted from the current \(J_{3}\) in Eq.~(\ref{eq:liu8}) as functions of the threshold parameter \(s_0\) and the Borel parameter \(M_B^2\).}
\label{fig:mass}
\end{center}
\end{figure*}

We next determine the continuum threshold \(s_0\) by examining the stability of the extracted mass with respect to \(M_B^2\). As shown in the left panel of Fig.~\ref{fig:mass}, the mass curves corresponding to different values of \(M_B^2\) intersect around $s_0 \simeq 29.0~\text{GeV}^2$, 
which gives the weakest dependence on \(M_B^2\) for the extracted mass~\cite{chenhuaxing:jiao-dian}. We therefore choose this result as the central value of the continuum threshold and take the working interval to be $28.0~\text{GeV}^2 \leq s_0 \leq 30.0~\text{GeV}^2$. 
Combining this interval with the lower and upper bounds on \(M_B^2\), we obtain the Borel window
$2.87~\text{GeV}^2 \leq M_B^2 \leq 3.29~\text{GeV}^2$ for the current \(J_3\).

\begin{figure*}[hbtp]
\begin{center}
\subfigure[]{\includegraphics[width=0.35\textwidth]{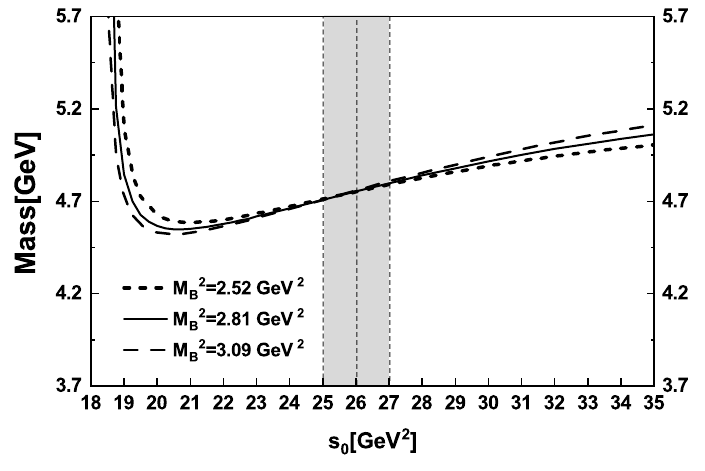}}
~~~~~~~~~~
\subfigure[]{\includegraphics[width=0.35\textwidth]{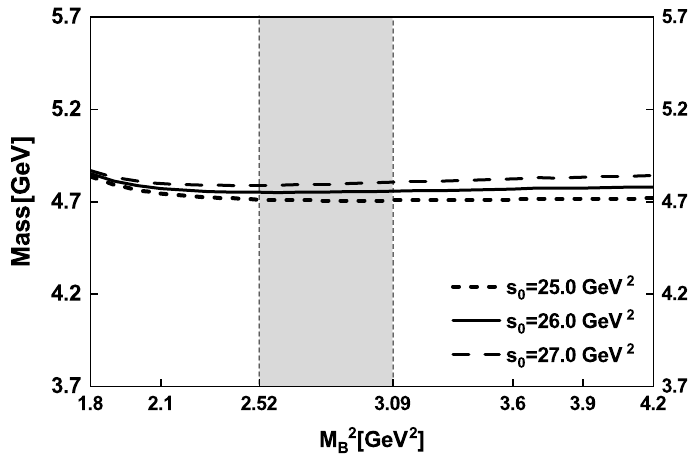}}
\caption{The mass \(M_{0^+}^{J_7}\), extracted from the current \(J_{7}\) in Eq.~(\ref{eq:liu5}), as functions of the threshold parameter \(s_0\) and the Borel parameter \(M_B^2\).}
\label{fig:mass5}
\end{center}
\end{figure*}

\begin{figure*}[hbtp]
\begin{center}
\subfigure[]{\includegraphics[width=0.35\textwidth]{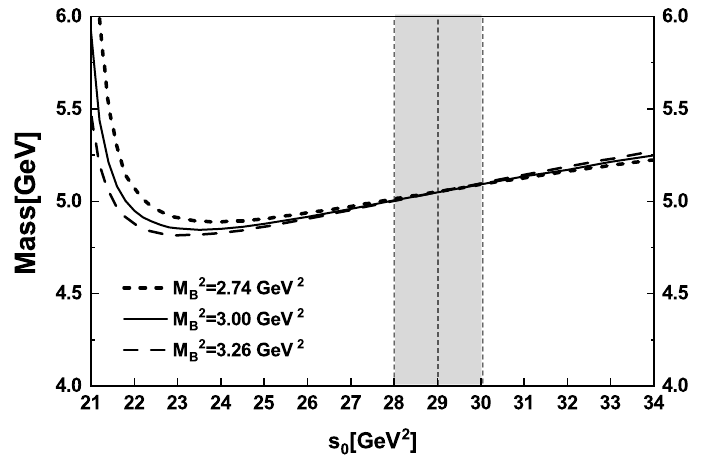}}
~~~~~~~~~~
\subfigure[]{\includegraphics[width=0.35\textwidth]{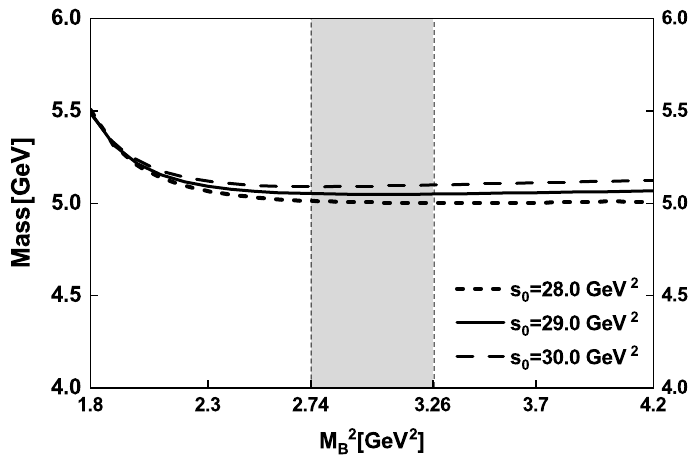}}
\caption{The mass \(M_{0^-}^{J_{15}}\), extracted from the current \(J_{15}\) in Eq.~(\ref{eq:liu11}), as functions of the threshold parameter \(s_0\) and the Borel parameter \(M_B^2\).}
\label{fig:mass11}
\end{center}
\end{figure*}

\begin{figure*}[hbtp]
\begin{center}
\subfigure[]{\includegraphics[width=0.35\textwidth]{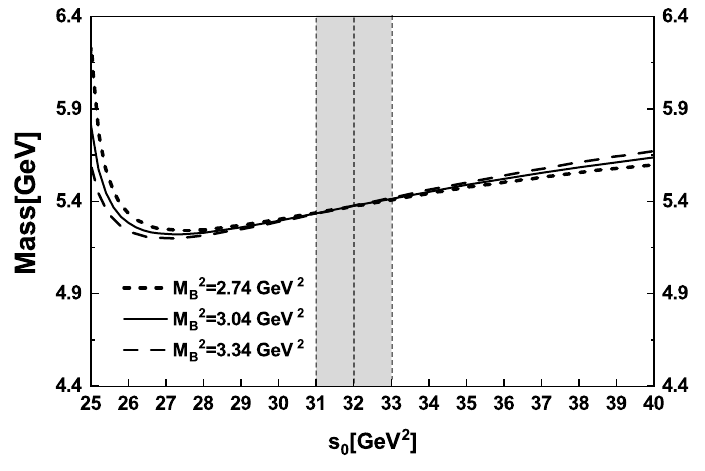}}
~~~~~~~~~~
\subfigure[]{\includegraphics[width=0.35\textwidth]{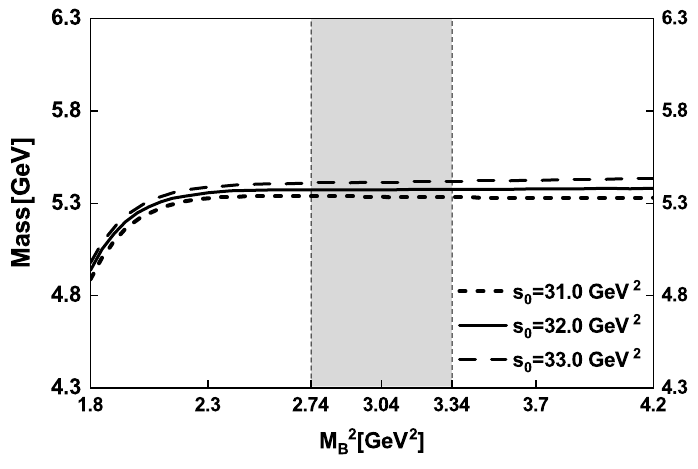}}
\caption{The mass \(M_{0^-}^{J_{18}}\), extracted from the current \(J_{18}\) in Eq.~(\ref{eq:liu16}), as functions of the threshold parameter \(s_0\) and the Borel parameter \(M_B^2\).}
\label{fig:mass16}
\end{center}
\end{figure*}

\begin{table*}[hpt]
\begin{center}
\renewcommand{\arraystretch}{1.25}
\caption{Extracted masses and current couplings for the \((c\bar{c})(c\bar{q})\) tetraquark candidates with \(J^P=0^{\pm}\).}
\begin{tabular}{c|c|c|c|c|c|c|c}
\hline\hline
~\multirow{2}{*}{Current}~& ~~\multirow{2}{*}{State}~~ & ~\multirow{2}{*}{~$s_0^{\rm min}~[{\rm GeV}^2]$~}~ & \multicolumn{2}{c|}{Working regions} & ~\multirow{2}{*}{Pole~[\%]}~ & ~\multirow{2}{*}{~Mass~[GeV]~}~&~\multirow{2}{*}{~Current coupling~[GeV$^5$]~}~
\\ \cline{4-5}
&&&~~$M_B^2~[{\rm GeV}^2]$~~&~$s_0~[{\rm GeV}^2]$~~&&&
\\ \hline\hline
$J_{3}$&$|(c\bar{c})(c\bar{q});0^+\rangle$&$26.7$&$2.87$-$3.29$&$29.0\pm1.0$&$40$-$52$&$5.00^{+0.12}_{-0.12}$&$0.205^{+0.028}_{-0.027}$ \\
$J_{4}$&$|(c\bar{c})(c\bar{q});0^+\rangle$&$27.5$&$3.13$-$3.50$&$29.0\pm1.0$&$40$-$49$&$5.04^{+0.12}_{-0.12}$&$0.110^{+0.018}_{-0.019}$\\
$J_{7}$&$|(c\bar{c})(c\bar{q});0^+\rangle$&$23.3$&$2.52$-$3.09$&$26.0\pm1.0$&$40$-$60$&$4.76^{+0.15}_{-0.15}$&$0.022^{+0.006}_{-0.006}$  \\
$J_{10}$&$|(c\bar{c})(c\bar{q});0^+\rangle$
&$26.6$&$2.58$-$3.04$&$28.5\pm1.0$&$40$-$52$&$4.98^{+0.12}_{-0.12}$&$0.029^{+0.007}_{-0.007}$
\\  \hline
$J_{15}$&$|(c\bar{c})(c\bar{q});0^-\rangle$&$26.8$&$2.74$-$3.26$&$29.0\pm1.0$&$40$-$55$&$5.04^{+0.15}_{-0.15}$&$0.018^{+0.004}_{-0.005}$\\
$J_{18}$&$|(c\bar{c})(c\bar{q});0^-\rangle$&$29.7$&$2.74$-$3.34$&$32.0\pm1.0$&$40$-$60$&$5.37^{+0.13}_{-0.13}$&$0.059^{+0.009}_{-0.009}$
\\ \hline\hline
\end{tabular}
\label{tab:results1}
\end{center}
\end{table*}

\begin{table*}[hpt]
\begin{center}
\renewcommand{\arraystretch}{1.25}
\caption{Extracted masses and current couplings for the \((b\bar{b})(b\bar{q})\) tetraquark candidates with \(J^P=0^{\pm}\).}
\begin{tabular}{c|c|c|c|c|c|c|c}
\hline\hline
~\multirow{2}{*}{Current}~& ~~\multirow{2}{*}{State}~~ & ~\multirow{2}{*}{~$s_0^{\rm min}~[{\rm GeV}^2]$~}~ & \multicolumn{2}{c|}{Working regions} & ~\multirow{2}{*}{Pole~[\%]}~ & ~\multirow{2}{*}{~Mass~[GeV]~}~&~\multirow{2}{*}{~Current coupling~[GeV$^5$]~}~
\\ \cline{4-5}
&&&~~$M_B^2~[{\rm GeV}^2]$~~&~$s_0~[{\rm GeV}^2]$~~&&&
\\ \hline\hline
$J_{3}$&$|(b\bar{b})(b\bar{q});0^+\rangle$&$196.5$&$8.84$-$10.97$&$206.5\pm1.0$&$40$-$59$&$13.90^{+0.10}_{-0.10}$&$2.825^{+0.132}_{-0.132}$ \\
$J_{4}$&$|(b\bar{b})(b\bar{q});0^+\rangle$&$197.1$&$9.09$-$11.25$&$207.0\pm1.0$&$40$-$59$&$13.91^{+0.10}_{-0.10}$&$1.440^{+0.068}_{-0.068}$\\
$J_{7}$&$|(b\bar{b})(b\bar{q});0^+\rangle$&$191.8$&$8.28$-$11.07$&$204.5\pm1.0$&$40$-$66$&$13.82^{+0.10}_{-0.10}$&$0.389^{+0.020}_{-0.020}$ \\
$J_{10}$&$|(b\bar{b})(b\bar{q});0^+\rangle$
&$195.8$&$8.33$-$10.66$&$207.5\pm1.0$&$40$-$63$&$13.92^{+0.10}_{-0.10}$&$0.456^{+0.023}_{-0.023}$
\\ \hline
$J_{15}$&$|(b\bar{b})(b\bar{q});0^-\rangle$&$202.4$&$8.89$-$10.21$&$209.0\pm1.0$&$40$-$53$&$14.03^{+0.10}_{-0.10}$&$0.160^{+0.007}_{-0.007}$ \\
$J_{18}$&$|(b\bar{b})(b\bar{q});0^-\rangle$&$207.0$&$8.75$-$9.75$&$211.5\pm1.0$&$40$-$50$&$14.06^{+0.16}_{-0.16}$&$0.430^{+0.023}_{-0.023}$
\\ \hline\hline
\end{tabular}
\label{tab:results2}
\end{center}
\end{table*}

\begin{table*}[hpt]
\centering
\renewcommand{\arraystretch}{1.15}
\begin{tabular}{|c|c|c|c|c|}
\hline
$J^{P}$ & Flavor content & Hadronic state & S-wave decay & P-wave decay \\
\hline
\multirow{4}{*}{$0^{+}$}
& \multirow{2}{*}{$(c\bar{c})(c\bar{q})$}
& $T_{3c,0}(4760)$ & -- & -- \\
\cline{3-5}
& & $T_{3c,0}(5000)$ & $\eta_{c}D$ & -- \\
\cline{2-5}
& \multirow{2}{*}{$(b\bar{b})(b\bar{q})$}
& $T_{3b,0}(13820)$ & -- & -- \\
\cline{3-5}
& & $T_{3b,0}(13910)$ & -- & -- \\
\hline
\multirow{3}{*}{$0^{-}$}
& \multirow{2}{*}{$(c\bar{c})(c\bar{q})$}
& $T_{3c,0}(5040)$ & -- & $\eta_{c}D^*,\, J/\psi D$ \\
\cline{3-5}
& & $T_{3c,0}(5370)$ & $\eta_{c}D^{*}_{0}(2300),\, \chi_{c0}(1P)D$ & $\eta_{c}D^*,\, J/\psi D$ \\
\cline{2-5}
& $(b\bar{b})(b\bar{q})$
& $T_{3b,0}(14030)$ & -- & -- \\
\hline
\end{tabular}
\caption{Possible two-body strong decay modes of the tetraquark candidates via the fall-apart mechanism. }
\label{tab:results3}
\end{table*}

It is also necessary to examine the dependence of the extracted mass on the parameters \(M_B^2\) and \(s_0\). In the left panel of Fig.~\ref{fig:mass}, the slope of the curve reflects the sensitivity of the extracted mass to \(s_0\): the smaller the slope, the weaker the dependence. In the adopted interval of \(s_0\), the slope remains moderate, indicating that the \(s_0\)-dependence is under control. It is worth noting that although there appears to be a stationary point around \(s_0 \sim 23.0~\text{GeV}^2\), no acceptable Borel window exists there because \(s_0 < s_0^{\rm min} = 26.7~\text{GeV}^2\). Therefore, this seeming stability point should be discarded. The right panel of Fig.~\ref{fig:mass} shows that the extracted mass depends only weakly on \(M_B^2\) within the window $2.87 \leq M_B^2 \leq 3.29~\text{GeV}^2$ for all the chosen $s_0$ values. 
These observations clearly validate the above working regions for $s_0$ and $M_B$.

With such preparations, for the current \(J_3\) we finally obtain 
\begin{eqnarray}
M_{0^+}^{J_3} &=& 5.00^{+0.12}_{-0.12}{\rm~GeV} \, ,
\label{eq:massliu8}
\\ \nonumber
f_{0^+}^{J_3} &=& 0.205^{+0.028}_{-0.027}{\rm~GeV^5} \, ,
\label{eq:decayliu8}
\end{eqnarray}
where the uncertainties originate from \(s_0\), \(M_B^2\), and the input parameters listed in Eq.~(\ref{eq:condensate}). We have also examined the violation of the factorization approximation for the four-gluon condensate by varying \(\kappa\) from 1 to 8~\cite{vacuum-saturation:1,vacuum-saturation:2}. The corresponding uncertainty turns out to be rather small and can be safely neglected, which is within expectation since the four-gluon condensate contributes only through higher-dimensional OPE terms with \(D\geq 8\). 

The same procedure can be applied to \(J_4\) and \(J_{10}\). As shown in Table~\ref{tab:results1}, the working windows of $s_0$ and $M_B$ are quite reasonable, and the extracted masses are
\begin{eqnarray}
 M_{0^+}^{J_4} &=& 5.04^{+0.12}_{-0.12}{\rm~GeV} \, ,
 \label{eq:massliu7}\\
M_{0^+}^{J_{10}} &=& 4.98^{+0.12}_{-0.12}{\rm~GeV} \, .
\label{eq:massliu4}
\end{eqnarray}
The masses extracted from \(J_3\), \(J_4\), and \(J_{10}\) are all around \(5.00\ \mathrm{GeV}\), which clearly suggest a \(0^+\) tetraquark candidate, denoted by \(T_{3c,0}(5000)\), with the mass 
\begin{eqnarray}
M_{0^+}^{J_{3,4,10}} &=& 5.00^{+0.16}_{-0.14}{\rm~GeV} \, .
\label{eq:mass-5000}
\end{eqnarray}

Although the interpolating currents \(J_3\), \(J_4\), and \(J_{10}\) have different color/Dirac structures, they yield very similar mass predictions within uncertainties. This likely indicates that these currents couple predominantly to the same lowest-lying \(0^+\) tetraquark state, rather than to distinct physical resonances. In QCD sum rules, different local currents carrying the same quantum numbers may have sizable overlap with the same hadronic state. A more definitive distinction would require an analysis of the off-diagonal correlation functions and possible current mixing.

For the current \(J_{7}\) in Eq.~(\ref{eq:liu5}), the working windows are found to be $2.52~\text{GeV}^2 \leq M_B^2 \leq 3.09~\text{GeV}^2$ and $25.0~\text{GeV}^2 \leq s_0 \leq 27.0~\text{GeV}^2$, 
as shown in Fig.~\ref{fig:mass5}. The extracted mass and current coupling are
\begin{eqnarray}
M_{0^+}^{J_7} &=& 4.76^{+0.15}_{-0.15}{\rm~GeV} \, ,
\label{eq:massliu5}
\\ \nonumber
f_{0^+}^{J_7} &=& 0.022^{+0.006}_{-0.006}{\rm~GeV^5} \, .
\label{eq:decayliu5}
\end{eqnarray}
Since the current \(J_7\) has a pseudoscalar--pseudoscalar and color-singlet--color-singlet structure, the extracted state, labeled as \(T_{3c,0}(4760)\), is naturally close to the \(\eta_c D\) threshold at $M_{\eta_c D} \approx 4.85~\text{GeV}$.
In fact, the mass of \(T_{3c,0}(4760)\) lies blow not only the \(\eta_c D\) threshold but also other relevant charmonium--open-charm thresholds. Therefore, none of the two-body strong decay channels via the fall-apart mechanism are open for this state. Its mass is compatible with a near-threshold \(\eta_c D\) configuration, and it is expected to be relatively narrow.

For the current \(J_{15}\) in Eq.~(\ref{eq:liu11}), the corresponding working regions are shown in Fig.~\ref{fig:mass11}. The resulting mass and coupling of the state from this current, labeled as \(T_{3c,0}(5040)\), are given by 
\begin{eqnarray}
M_{0^-}^{J_{15}} &=& 5.04^{+0.15}_{-0.15}{\rm~GeV} \, ,
\label{eq:massliu11}
\\ \nonumber
f_{0^-}^{J_{15}} &=& 0.018^{+0.004}_{-0.005}{\rm~GeV^5} \, .
\label{eq:decayliu11}
\end{eqnarray}
Owing to the color-singlet-color-singlet structure, the \(T_{3c,0}(5040)\) can decay into \(\eta_c D^*\) and \(J/\psi D\) via P-wave channels, which are allowed both kinematically and by parity quantum numbers.

For the current \(J_{18}\) in Eq.~(\ref{eq:liu16}), the working regions are determined to be $2.74~\text{GeV}^2 \leq M_B^2 \leq 3.34~\text{GeV}^2$ and $31.0~\text{GeV}^2 \leq s_0 \leq 33.0~\text{GeV}^2$. 
The dependence of the extracted mass on \(M_B^2\) and \(s_0\) is shown in Fig.~\ref{fig:mass16}, and the resulting mass and current coupling are 
\begin{eqnarray}
M_{0^-}^{J_{18}} &=& 5.37^{+0.13}_{-0.13}{\rm~GeV} \, ,
\label{eq:massliu16}
\\ \nonumber
f_{0^-}^{J_{18}}&=& 0.059^{+0.009}_{-0.009}{\rm~GeV^5} \, .
\label{eq:decayliu16}
\end{eqnarray}
The hadronic state extracted from the current \(J_{18}\) is denoted as \(T_{3c,0}(5370)\). 
The mass of the corresponding tetraquark candidate \(T_{3c,0}(5370)\) lies above several charmonium--open-charm thresholds. Therefore, it may decay through the S-wave channels \(\eta_c D^{*}_{0}(2300)\) and \(\chi_{c0}(1P) D\), as well as the P-wave channels \(\eta_c D^{*}\) and \(J/\psi D\).

We then extend the studies to the \((b\bar{b})(b\bar{q})\) systems, and the corresponding results are listed in Table~\ref{tab:results2}. We find two different states with \(J^P=0^+\) at masses around 13.82 and 13.91~GeV, which are labeled as \(T_{3b,0}(13820)\) and \(T_{3b,0}(13910)\), respectively. Qualitatively similar states are also reported for \(bb\bar{b}\bar{q}\) tetraquark systems in Ref.~\cite{triply:QCDsumrules-1}. For the \(J^P=0^-\) bottom sector, the state extracted from \(J_{18}\) carries a somewhat larger uncertainty than that from \(J_{15}\), but remains compatible with the same overall mass region. The state extracted from \(J_{15}\) turns out to be more stable and can be identified with \(T_{3b,0}(14030)\). Taking the quoted uncertainties into account, the masses of the \((b\bar{b})(b\bar{q})\) states are expected to lie in the ranges \(13.72\text{--}14.02\) GeV for \(J^P=0^+\) and \(13.90\text{--}14.22\) GeV for \(J^P=0^-\).  

The possible two-body strong decay modes via the fall-apart mechanism are summarized in Table~\ref{tab:results3}. The states \(T_{3c,0}(5000)\), \(T_{3c,0}(5040)\), and \(T_{3c,0}(5370)\) all lie above at least one relevant charmonium--open-charm threshold and can therefore decay through the fall-apart mechanism. Since the available phase space is limited, they may still be relatively narrow. In contrast, \(T_{3c,0}(4760)\) and all the predicted \((b\bar{b})(b\bar{q})\) states lie below the corresponding bottomonium-- or charmonium--heavy-meson thresholds relevant to such two-body strong decays. Therefore, the corresponding two-body strong decays via the fall-apart mechanism are kinematically forbidden for these states. They may still decay through heavy-quark annihilation, electromagnetic transitions, or weak interactions, but such kinds of decay widths are expected to be comparatively small.

\section{Conclusion}
\label{sec:summary}

Within the framework of QCD sum rules, we have systematically studied the mass spectra and possible decay patterns of the \((c\bar{c})(c\bar{q})\) and \((b\bar{b})(b\bar{q})\) tetraquark systems with quantum numbers \(J^{P}=0^{\pm}\). Using two different color structures, \( [8_c]_{Q\bar{Q}} \otimes [8_c]_{Q\bar{q}} \) and \( [1_c]_{Q\bar{Q}} \otimes [1_c]_{Q\bar{q}} \), we construct 18 interpolating currents and derive the corresponding QCD sum rules from the two-point correlation functions by including the OPE contributions up to dimension 9. 

For the \((c\bar{c})(c\bar{q})\) system, our numerical analysis suggests four possible states: two \(0^+\) states, \(T_{3c,0}(4760)\) and \(T_{3c,0}(5000)\); and two \(0^-\) states, \(T_{3c,0}(5040)\) and \(T_{3c,0}(5370)\). For the \((b\bar{b})(b\bar{q})\) system, the extracted masses lie in the ranges \(13.77\text{--}13.97\) GeV for the \(0^+\) states and \(13.95\text{--}14.17\) GeV for the \(0^-\) states after including the uncertainties. The \(0^-\) result from the current \(J_{18}\) has a somewhat larger uncertainty than that from \(J_{15}\), but the two currents lead to states with compatible masses.

As shown in Tables~\ref{tab:results1}--\ref{tab:results3}, the states \(T_{3c,0}(5000)\), \(T_{3c,0}(5040)\), and \(T_{3c,0}(5370)\) lie above at least one relevant charmonium--open-charm threshold, and thus the corresponding two-body strong decays via the fall-apart mechanism are kinematically allowed. In contrast, \(T_{3c,0}(4760)\) lies below the relevant charmonium plus \(D^{(*)}\) thresholds, and all predicted \((b\bar{b})(b\bar{q})\) states in this work lie below the corresponding bottomonium plus \(\bar{B}^{(*)}\) thresholds. Therefore, the corresponding two-body strong decays via the fall-apart mechanism are kinematically forbidden for these states. It is worth noting that some \(bb\bar{b}\bar{q}\) states are also found to lie below the bottomonium plus \(\bar{B}^{(*)}\) thresholds and may be relatively narrow~\cite{triply:QCDsumrules-1}.

Nevertheless, even when the fall-apart channels are closed, other decay mechanisms may still exist. In particular, heavy-quark pair annihilation followed by light-quark pair creation $(Q\bar{Q})(Q\bar{q}) \to (Q\bar{q}) + (q\bar{q})$, 
can in principle lead to final states containing a \(D^{(*)}\) or \(\bar{B}^{(*)}\) meson together with light hadrons. Electromagnetic decays are also possible and may generate final states with an additional photon. Therefore, \(T_{3c,0}(4760)\) and the predicted \((b\bar{b})(b\bar{q})\) states should not be regarded as absolutely stable; rather, they are stable only with respect to the two-body strong decays via the fall-apart mechanism discussed above, and are expected to be comparatively narrow.

In summary, the present QCD sum rule analysis supports the possible existence of several triply heavy tetraquark candidates in the \((c\bar{c})(c\bar{q})\) and \((b\bar{b})(b\bar{q})\) systems with \(J^{P}=0^{\pm}\), including both states above the relevant fall-apart thresholds and states lying below them that may therefore be comparatively narrow. This pattern provides a useful guide for future experimental studies, since the above-threshold states are expected to decay more readily through strong interactions, while the below-threshold states may manifest themselves as narrower structures in appropriate invariant-mass distributions. With the high luminosity available at the LHC and the continuously growing data sample at Belle II, experimental searches for such triply heavy tetraquark configurations in heavy-flavor final states should become increasingly feasible, which will provide an important test of the present predictions and, more generally, deepen our understanding of multiquark dynamics in the heavy-quark sector.

\section*{Acknowledgments}

This work is supported in part by Hebei Natural Science Foundation under Grants No.~A2026205016, and by the Science Foundation of Hebei Normal University with Contract No.~L2023B09. 

\bibliographystyle{elsarticle-num}
\bibliography{ref}

\newpage
\begin{widetext}
\appendix
\section{The OPE spectral density \texorpdfstring{$\rho_{\rm OPE}(s)$}{rho\_OPE(s)}}
\label{sec:appendix}

In the Appendix, we give the analytical expression for the OPE spectral density only for one representative case, while the results for the other currents are omitted for brevity. As an explicit example, we present the spectral density corresponding to the current \(J_{3}^{\bar{c}c\bar{c}q}\). We define
\[
\tilde{m}_c^2 \equiv m_c^2 \left( \frac{1}{x} + \frac{1}{y} + \frac{1}{1-x-y} \right).
\]
The OPE spectral density derived from the current \(J_3\) is written as
\begin{equation}
    \rho_{3}(s) = \rho^{(0)}_{3}(s) + \rho^{(3)}_{3}(s) + \rho^{(4)}_{3}(s) + \rho^{(5)}_{3}(s)+ \rho^{(6)}_{3}(s) + \rho^{(7)}_{3}(s) + \rho^{(8)}_{3}(s) + \rho^{(9)}_{3}(s) \, , \tag{A.1}
\end{equation}
where the superscript \((n)\) denotes the contribution of the vacuum condensates of dimension \(n\), with \(n=0,3,4,5,6,7,8,9\). For the sake of compactness, some higher dimensional terms proportional to \(\delta(s-\tilde{m}_c^2)\) are written directly in the form used in the Borel sum rules, and therefore explicit \(M_B^2\) dependence appears in those contributions. The explicit expressions are given below:
\begin{equation*}
\begin{aligned}
    \rho^{(0)}_{3}(s)&=
\int^{x_{max}}_{x_{min}}dx\int^{y_{max}}_{y_{min}}dy \,\Bigg\{ -\frac{\left(s x y (x + y - 1) - m_c^2 (x^2 + x (y - 1) + (y - 1) y)\right)^4}{8 \pi^6 x^3 y^3 (x + y - 1)^3}
 \Bigg\} ,
\end{aligned}
\end{equation*}
\begin{equation*}
\begin{aligned}
    \rho^{(3)}_{3}(s)&=\int^{x_{max}}_{x_{min}}dx\int^{y_{max}}_{y_{min}}dy \,\Bigg\{ \frac{4 \langle \bar{q}q\rangle m_c^3 \left( \frac{m_c^2 (x^2 + x (y - 1) + (y - 1) y)}{x y (x + y - 1)} - s \right)}{\pi^4}\Bigg\} ,
\end{aligned}
\end{equation*}
\begin{equation*}
\begin{aligned}
\rho^{(4)}_{3}(s)&=\int^{x_{max}}_{x_{min}}dx\int^{y_{max}}_{y_{min}}dy \,\Bigg\{
-\frac{\langle \alpha_s G G \rangle}{96 \pi^5 x^3 y^3 (x + y - 1)^3} \bigg[ -2 s x y m_c^2 \bigg( 4 x^7 (3 y + 2) + x^6 (87 y^2 - 29 y - 32) \\
&+ x^5 (243 y^3 - 309 y^2 + 14 y + 48)  + 2 x^4 (183 y^4 - 385 y^3 + 212 y^2 + 6 y - 16) \\
&+ x^3 (333 y^5 - 910 y^4 + 847 y^3 - 268 y^2 - 10 y + 8) \quad + x^2 (y - 1)^2 y (177 y^3 - 205 y^2 + 67 y + 1) \\
& + x (y - 1)^3 y^2 (42 y^2 - 13 y - 1) + 8 (y - 1)^4 y^3 \bigg) + m_c^4 \bigg( 8 x^8 (y + 2) + x^7 (58 y^2 + 19 y - 64) \\
&+ 2 x^6 (85 y^3 - 74 y^2 - 61 y + 48) + 2 x^5 (147 y^4 - 265 y^3 + 64 y^2 + 86 y - 32) \\
&+ 2 x^4 (167 y^5 - 405 y^4 + 298 y^3 - 21 y^2 - 47 y + 8) + x^3 (y - 1)^2 y (250 y^3 - 210 y^2 + 36 y + 17) \\
&+ 2 x^2 (y - 1)^2 y^2 (59 y^3 - 46 y^2 + 8 y + 1) + x (y - 1)^3 y^3 (28 y^2 + 33 y - 17) + 16 (y - 1)^4 y^4 \bigg) \\
&+ s^2 x^3 y^3 (x + y - 1)^3 (16 x^2 + x (68 y - 29) + 56 y^2 - 59 y - 1) \bigg]
 \Bigg\},
\end{aligned}
\end{equation*}

\begin{equation*}
\begin{aligned}
\rho^{(5)}_{3}(s)&=\int^{x_{max}}_{x_{min}}dx\int^{y_{max}}_{y_{min}}dy \,\Bigg\{ \frac{\langle g_s \bar{q} \sigma G q \rangle m_c^3 \left(24  (x + y - 1) + 1\right)}{12 \pi^4 (x + y - 1)}\Bigg\}\\
&+\int^{1}_{0}dx\int^{1-x}_{0}dy \,\Bigg\{\frac{\langle g_s \bar{q} \sigma G q \rangle \tilde{m}_c^2 m_c^3 \delta\left(s - \tilde{m}_c^2\right)}{\pi^4}\Bigg\},
\end{aligned}
\end{equation*}

\begin{equation*}
\begin{aligned}
\rho^{(6)}_{3}(s)&=\int^{x_{max}}_{x_{min}}dx\int^{y_{max}}_{y_{min}}dy \,\Bigg\{ \frac{\langle g_s^3 G^3 \rangle}{2304 \pi^6 x^3 y^3 (x+y-1)^3} \Bigg[ m_c^2 \bigg(
8 x^8 (5 y - 9) + x^7 (160 y^2 - 513 y + 288)\\
&+ x^6 (354 y^3 - 1105 y^2 + 1318 y - 432) + x^5 (632 y^4 - 1481 y^3 + 1863 y^2 - 1276 y + 288) \\
&+ x^4 (932 y^5 - 2086 y^4 + 1709 y^3 - 1045 y^2 + 450 y - 72) + x^3 y (954 y^5 - 2661 y^4 + 2319 y^3 - 714 y^2 + 121 y - 19) \\
&+ x^2 (y - 1)^2 y^2 (560 y^3 - 1125 y^2 + 163 y + 6)
+ x (y - 1)^3 y^3 (140 y^2 - 443 y + 19)
- 72 (y - 1)^4 y^4
\bigg)\\
&- 3 s x y \bigg(
4 x^7 (3 y - 2)
+ x^6 (48 y^2 - 93 y + 32)
+ x^5 (117 y^3 - 215 y^2 + 206 y - 48) \\
&+ 2 x^4 (111 y^4 - 177 y^3 + 140 y^2 - 90 y + 16)
+ x^3 (267 y^5 - 584 y^4 + 377 y^3 - 106 y^2 + 54 y - 8) \\
&+ x^2 (y - 1)^2 y (168 y^3 - 219 y^2 - 6 y + 1)
+ x (y - 1)^3 y^2 (42 y^2 - 77 y - 1)
- 8 (y - 1)^4 y^3
\bigg)
\Bigg]\Bigg\}\\
&+\int^{1}_{0}dx\int^{1-x}_{0}dy \,\Bigg\{\frac{\langle g_s^3 G^3 \rangle m_c^2 \tilde{m}_c^2  \delta\left(s - \tilde{m}_c^2\right)}{1152 \pi^6 x^2 y^2 (x + y - 1)^2} \bigg(4 x^6 + 12 x^5 (y - 1) + 12 x^4 (y - 1)^2 \\
&+ x^3 (17 y^3 - 12 y^2 + 12 y - 4) + 42 x^2 (y - 1) y^3 + 42 x (y - 1)^2 y^3 + 14 (y - 1)^3 y^3\bigg)\Bigg\},
\end{aligned}
\end{equation*}

\begin{equation*}
\begin{aligned}
\rho^{(7)}_{3}(s)&=\int^{1}_{0}dx\int^{1-x}_{0}dy \,\Bigg\{\frac{\langle \alpha_s G G \rangle \langle \bar{q}q\rangle m_c^3 \,
\delta\!\left(s - \tilde{m}_c^2\right)}{18 \pi^3 M_B^2 x^3 y^3 (x+y-1)^3} \Bigg[4 m_c^2 \Big(
x^6
+ 3 x^5 (y - 1)
+ 3 x^4 (y - 1)^2
+ x^3 (y - 1)^3 \\
&+ 3 x^2 (y - 1) y^3
+ 3 x (y - 1)^2 y^3
+ (y - 1)^3 y^3
\Big)+ M_B^2 x y \Big(
- 12 x^5
+ x^4 (36 - 23 y) \\
&+ x^3 (-31 y^2 + 47 y - 36)
+ x^2 (-41 y^3 + 54 y^2 - 25 y + 12)
- x (y - 1)^2 y (33 y - 1)
- 12 (y - 1)^3 y^2
\Big)
\Bigg]\Bigg\},
\end{aligned}
\end{equation*}

\begin{equation*}
\begin{aligned}
\rho^{(8)}_{3}(s)&=\int^{x_{max}}_{x_{min}}dx\int^{y_{max}}_{y_{min}}dy \,\Bigg\{ \frac{\langle \alpha_s G G \rangle^2}{62208 \pi ^4 x^2 y^2 (x+y-1)^2}\Bigg[ 480 x^6 + 4 x^5 (-555 + 406 y + 12 y^2) \\
&+ 60 (-1 + y)^2 y^2 (-1 - 31 y + 28 y^2) + x^4 (2940 - 4959 y + 1918 y^2 + 264 y^3) \\
&+ 3 x^3 (-380 + 1117 y - 1397 y^2 + 796 y^3 + 128 y^4) + x y (-15 - 31 y + 5871 y^2 - 11109 y^3 + 5284 y^4) \\
&+  x^2 (-60 - y + 636 y^2 - 6561 y^3 + 5338 y^4 + 168 y^5)\Bigg]\Bigg\}+\int^{1}_{0}dx\int^{1-x}_{0}dy \,\Bigg\{\frac{\langle \alpha_s G G \rangle^2 m_c^2 \delta \left(s-\tilde{m}_c^2\right)}{186624 \pi ^4 x^4 y^4 M_B^2 (x+y-1)^4}\\
&\times\Bigg[ x y M_B^2\left(72 x^9-2 x^8 \left(48 y^2+94 y+1467\right) \right. -x^5 \left(3024 y^5+13479 y^4-32110 y^3+501 y^2+19114 y-9468\right)\\
&-x^7 \left(720 y^3+2065 y^2+8213 y-10818\right)-x^6 \left(2016 y^4+6856 y^3+1526 y^2-23479 y+15372\right)\\
&+x^4 \left(-2736 y^6-14979 y^5+65726 y^4-63857 y^3+16282 y^2+1418 y-1854\right)\\
&-x^3 \left(1440 y^7+8056 y^6-55540 y^5+92777 y^4-58369 y^3+14221 y^2-2783 y+198\right)\\
&-x^2 (y-1)^2 y \left(336 y^5+1747 y^4-14966 y^3+19432 y^2-1962 y+165\right)\left.+x (y-1)^3 y^2 \left(682 y^3+163 y^2-4178 y+165\right) \right.\\
&\left.+18 (y-1)^4 y^3 \left(14 y^2-47 y-11\right)\right)+2 m_c^2 \left(396 x^{10}-4 x^9 \left(3 y^2-461 y+495\right)+x^8 \left(-90 y^3+3886 y^2-7744 y+3960\right) \right.\\
&+x^7 \left(-264 y^4+4992 y^3-12475 y^2+12536 y-3960\right)+x^6 \left(-456 y^5+5807 y^4-11437 y^3+15055 y^2-9584 y+1980\right)\\
&+x^5 \left(-516 y^6+9587 y^5-13775 y^4+9910 y^3-8126 y^2+3316 y-396\right)\\
&+x^4 y \left(-384 y^6+16307 y^5-30365 y^4+17326 y^3-3276 y^2+1624 y-368\right)\\
&+x^3 y^2 \left(-180 y^6+19482 y^5-50467 y^4+44920 y^3-13686 y^2-122 y+53\right)\\
&-x^2 (y-1)^2 y^2 \left(42 y^5-14962 y^4+20939 y^3-6875 y^2-13 y+5\right)\left.+4 x (y-1)^4 y^4 (1706 y-347)+1386 (y-1)^5 y^5\right) \Bigg]\Bigg\},
\end{aligned}
\end{equation*}

\begin{equation*}
\begin{aligned}
\rho^{(9)}_{3}(s)&=\int^{1}_{0}dx\int^{1-x}_{0}dy \,\Bigg\{-\frac{\langle g_s^3 G^3 \rangle \langle \bar{q}q\rangle m_c^3 \delta \left(s-\tilde{m}_c^2\right)}{36 \pi ^4 x^4 y^4 M_B^4 (x+y-1)^4}\Bigg[ m_c^2 \left(x^8+4 x^7 (y-1)+6 x^6 (y-1)^2+4 x^5 (y-1)^3 \right.  \\
&\left.+x^4 \left(3 y^4-4 y^3+6 y^2-4 y+1\right)+4 x^3 (y-1) y^4+6 x^2 (y-1)^2 y^4+4 x (y-1)^3 y^4+(y-1)^4 y^4\right)\\
&-3 x y M_B^2 \left(x^7+4 x^6 (y-1)+6 x^5 (y-1)^2+4 x^4 (y-1)^3+x^3 \left(4 y^4-7 y^3+6 y^2-4 y+1\right)\right.\\
&\left.+6 x^2 (y-1)^2 y^3+4 x (y-1)^3 y^3+(y-1)^4 y^3\right)\Bigg]\\
&+\frac{\langle \alpha_s G G \rangle \langle g_s \bar{q} \sigma G q \rangle m_c^3 \delta \left(s-\tilde{m}_c^2\right)}{864 \pi ^3 x^4 y^4 M_B^6 (x+y-1)^4}\Bigg[ 4 x y M_B^2 m_c^2 \left(48 x^7+x^6 (153 y-193)+3 x^5 \left(90 y^2-167 y+97\right) \right.\\
&\left.+3 x^4 \left(111 y^3-240 y^2+194 y-65\right)+x^3 \left(363 y^4-806 y^3+666 y^2-273 y+49\right) \right.\\
&\left.+3 x^2 y \left(110 y^4-280 y^3+242 y^2-85 y+13\right)+3 x (y-1)^2 y^2 \left(61 y^2-75 y+13\right)+(y-1)^3 y^3 (48 y-49)\right)\\
&+x^2 y^2 M_B^4 (x+y-1)^3 \left(12 x^2+x y+12 y^2\right)-48 m_c^4 \left(x^8+4 x^7 (y-1)+x^6 \left(7 y^2-13 y+6\right) \right.\\
&+x^5 (y-1)^2 (7 y-4)+x^4 \left(7 y^4-16 y^3+15 y^2-7 y+1\right)+x^3 (y-1)^2 y \left(7 y^2-2 y+1\right)\\
&\left.+x^2 (y-1)^2 y^3 (7 y-4)+x (y-1)^3 y^3 (4 y-1)+(y-1)^4 y^4\right) \Bigg]\Bigg\}\,.
\end{aligned}
\end{equation*}

The integration limits appearing in the above equations are given by
\[
x_{\rm min} \equiv \frac{-3 m_c^2 + s - \sqrt{9 m_c^4 - 10 m_c^2 s + s^2}}{2s}, \qquad
x_{\rm max} \equiv \frac{-3 m_c^2 + s + \sqrt{9 m_c^4 - 10 m_c^2 s + s^2}}{2s},
\]
\[
y_{\rm min} \equiv \frac{1 - x - \sqrt{4 (m_c^2 x - m_c^2 x^2)/ (s x - m_c^2) + (1 - x)^2}}{2}, \qquad
y_{\rm max} \equiv \frac{1 - x + \sqrt{4 (m_c^2 x - m_c^2 x^2)/ (s x - m_c^2) + (1 - x)^2}}{2}.
\]

\end{widetext}

\end{document}